\documentclass[11pt]{amsart}
\allowdisplaybreaks[4]
\usepackage{amssymb}
\usepackage{amsfonts}
\usepackage{graphicx}
\usepackage{epstopdf}
\usepackage{dcolumn}
\usepackage{amsmath}
\usepackage{enumerate}
\usepackage{latexsym,bm}
\usepackage{slashed}
\usepackage{float}
\usepackage{cite}
\usepackage{CJK}
\usepackage[all,cmtip]{xy}
\usepackage{appendix}
\usepackage[colorlinks,
linkcolor=black,
citecolor=red
]{hyperref}

\DeclareMathOperator{\Tr}{Tr}

\DeclareMathOperator{\tr}{tr}
\DeclareMathOperator{\str}{str}
\DeclareMathOperator{\w}{wt}

\DeclareMathOperator{\Id}{Id}
\DeclareMathOperator{\re}{Re}

\theoremstyle{plain} \newtheorem{Cor}{Corollary}[section]
 \newtheorem{Def}[equation]{Definiton}
\newtheorem{Thm}[equation]{Theorem} \newtheorem{lem}[equation]{Lemma}
\newtheorem{Ex}[equation]{Example} \newtheorem{prop}[equation]{Proposition}  
\numberwithin{equation}{section}

\def\bd{\begin{Def}}
\def\ed{\end{Def}}
\def\bex{\begin{Ex}}
\def\eex{\end{Ex}}
\def\bpr{\begin{prop}}
\def\epr{\end{prop}}
\def\bp{\begin{proof}}
\def\ep{\end{proof}}
\def\bl{\begin{lem}}
\def\el{\end{lem}}
\def\bc{\begin{Cor}}
\def\ec{\end{Cor}}
\def\bt{\begin{Thm}}
\def\et{\end{Thm}}
\def \be {\begin{equation}}
\def \ee {\end{equation}}
\def\H{\mathcal{H}}

\def\C{\mathbb{C}}

\def\O{\Omega}
\def\a{\alpha}
\def\b{\beta}
\def\p{\partial}
\def\bap{\bar{\partial}}
\def\bj{\bar{j}}
\def\m{\mathcal}

\def\l{\lambda}

\begin{document}

\title{$tt^*$ Geometry, Singularity Torsion and Anomaly Formulas}
\author{Xinxing Tang}
\address{\noindent
BICMR,
Peking University, Beijing, China}
\email{xxtang@pku.edu.cn}
\maketitle

\begin{abstract}
This paper is concerned with the Schr\"odinger operators $\Delta_{f_0}$ and $\Delta_f$ attached to a pair $(\mathbb{C}^n, f_0)$ and its deformation $(\mathbb{C}^n, f)$, where $f_0$ is a non-degenerate and quasi-homogeneous polynomial on $\mathbb{C}^n$ and $f$ is its relevant or marginal deformation. We give the $tt^*$ geometry structure on the Hodge bundle associated to $\Delta_f$, which describes the genus 0 anomaly. Next we study the corresponding singularity torsion type invariants and give the anomaly formulas for the 2nd torsion type invariant.
\end{abstract}

\tableofcontents

\section{Introduction}

The torsion type invariant of this paper has its origins in \cite{Cecotti1993Ising} in the physical literature, where Cecotti and Vafa considered the K\"{a}hler metric in the $tt^*$ geometry, which is given by
$$ds^2=K_{i\bar{j}}dt^id\bar{t}^j\quad \text{ with }\quad K_{i\bar{j}}=<\phi_i\bar{\phi}_{\bj}>_{\text{torus}}.$$
They considered a torus with the fields $\phi_i$ and $\bar{\phi}_{\bj}$ inserted on the left and right side of
a flat torus respectively which are infinitely separated by two long tubes each with perimeter $1$. More precisely,
$$K_{i\bar{j}}=\lim_{L\rightarrow \infty} \Tr[(-1)^F\phi_ie^{-LH}\bar{\phi}_{\bj}e^{-LH}]=\tr(C_i\bar{C}_{\bj}).$$

They also considered a path integral of the form
\begin{equation}\label{K}
K=-4\int_{\mathcal{F}}\frac{d^2\rho}{\rho_{2}}\Tr[(-1)^{F}F_{L}F_{R}q^{H_L}\bar{q}^{H_R}],
\end{equation}
where $\rho=\rho_1+i\rho_2$ and $\mathcal{F}$ is the standard fundamental domain for $\mbox{SL}(2,\mathbb{Z})$, and  showed that $K$ served as the K\"{a}her potential of $K_{i\bar{j}}$ through the following identity
$$\p_i\bap_{\bj}K=\tr(C_i\bar{C}_{\bj})\footnote{The formula here ignores the contact term, which is given by $-\frac{G_{i\bar{j}}}{12}\Tr(-1)^F$; see in \cite{Bershadsky1993Holomorphic}.}.$$

In \cite{Cecotti1993Ising}, Cecotti and Vafa also commented that if the central charge $\hat{c}$ has the right value, \eqref{K} gives the one-loop correction to the gravitational coupling for a type II superstring compactified on the given N=2 superconformal model. In this case, $K$ could be viewed as a counterpart of genus 1 partition function $F_1$ in the CY B-model via the LG/CY correspondence.

The CY B-model is related to the deformation theory of complex structures. The genus 0 theory could be understood as the variation of Hodge structures. The higher genus information is more interesting and difficult. In the physics literature, Bershadshy-Cecotti-Ooguri-Vafa \cite{Bershadsky1994Kodaira} proposed a so-called Kodaira-Spencer gauge theory of gravity on Calabi-Yau 3-folds. BCOV suggested that the genus expansion of B-model should correspond to the Feynman diagram expansion of the gauge theory.
The genus 1 partition function is given by
\begin{equation}\label{BCOVF1}
F_1=\frac{1}{2}\int_{\mathcal{F}}\frac{d^2\tau}{\tau_2}\Tr[(-1)^FF_LF_Rq^{H_L}\bar{q}^{H_R}],
\end{equation}
which admits a holomorphic anomaly equation
$$\p_i\bap_{\bj}F_1=\frac{1}{2}\tr C_i\bar{C}_{\bj}-\frac{G_{i\bar{j}}}{24}\Tr(-1)^F.\footnote{In this paper, we use ``$\Tr$" to denote the trace on the infinite dimensional Hilbert space(cochain level), and ``$\tr$" to denote the trace on the cohomology level.}$$
When they considered the 1d CY B-model, they found that $F_1$ has a relation with the Ray-Singer torsion \cite{Ray1973Analytic}. That is,
$$F_1=\frac{1}{2}\sum_q(-1)^q\log I(\wedge^q T^*),$$
where $\log I(\wedge^q T^*)$ is the Ray-Singer torsion on the holomorphic bundle $\wedge^q T^*$. Therefore, the heat kernel theory entered the stage. Recall that Bismut, Gillet, Soule \cite{SouleAnalytic} extended the definition to the analytic torsion forms using Quillen's superconnectoin and developed the holomorphic anomaly equation of analytic torsion forms.
In the mathematical literature, the genus 1 invariant was defined rigorously by H. Fang, Z.Lu and K.Yoshikawa \cite{Fang2008Asymptotic}.
In \cite{Costello2012Quantum}, Costello and S. Li initiated a mathematical analysis of the generalized BCOV theory. They made a good use of the heat kernel to define the propagator of the Feynman diagrams.

Nonlinear Sigma B-model has another important part, which is called the LG B-model; it is related to the deformation of a singularity. Its genus 0 theory is given by Saito's theory of primitive forms \cite{Saito1981Primitive} and higher residue pairing \cite{Saito1983The}. In \cite{LiquantumB}, S. Li developed the general framework of perturbative quantization of the LG-twisted BCOV theory.

The LG B-model is closely related to the CY B-model. For example, for $(\mathbb{C}^5, f(\textbf{z})=z_1^5+\cdots+z_5^5)$, the corresponding CY 3-fold $M$ is given by $M=\{[\textbf{z}]|f(\textbf{z})=0\}\subset\mathbb{P}^4$, and the quantum information of $M$ can be read form $(\mathbb{C}^5, f)$. Such a phenomenon is called the LG/CY correspondence. Influenced by this correspondence, Fan and Fang \cite{Fan2016Torsion} started from the 1d LG B-model, i.e. supersymmetric quantum mechanics (SQM), and defined a counterpart of the BCOV torsion. Given a pair $(\mathbb{C}^n,f_0)$, where $f_0$ is a non-degenerate and quasi-homogeneous polynomial on $\mathbb{C}^n$, there is a 2nd-order operator $\Delta_{f_0}$ in the Schr\"{o}dinger representation of SQM. Under a certain assumption, they proceeded the heat kernel analysis of $e^{-t\Delta_{f_0}}$
and proved that $e^{-t\Delta_{f_0}}$ is trace-class. They defined the $i$-th zeta function as follows:
$$\zeta_{f_0}^i(s)=\frac{1}{2}\sum_{k=0}^{2k}(-1)^{k}k^i\frac{1}{\Gamma(s)}\int_0^{\infty}t^{s-1}\Tr(e^{-t\Delta_{f_0}^k}-\Pi_{0k})dt,$$
where $\Delta_{f_0}^k$ means restricting $\Delta_{f_0}$ to the $k$-forms, $\Pi_{0k}$ is the projection to the $k$-harmonic forms. Furthermore, the $i$-th torsion type invariants $T^i(f_0)$ are defined via the regularized $i$-th zeta function $\zeta_{f_0}^{i,R}(s)$ by
$$\log T^i(f_0)=-(\zeta_{f_0}^{i,R})'(0).$$
They studied the $i$-th torsion type invariants through heat kernel analysis of $\Delta_{f_0}$, and proved that the first torsion type invariant $T^1(f_0)$ vanishes and the second torsion type invariant $T^2(f_0)$ satisfies some summation property. We will further recall their results in Sections 4 and 6.

In this paper, we study the deformation counterpart of Fan-Fang's story. First, we start from the deformation $f$ of $f_0$ and give the $tt^*$ geometry structure on the deformation space which is well-known in \cite{Cecotti1991Topological}, \cite{Fan2011Schr}. The $tt^*$ equations could be viewed as the genus 0 anomaly equations in terms of \cite{Cecotti1992A}. Next, we continue the discussion of heat kernel analysis for $\Delta_f$ following from Fan-Fang's work. We first solve the existence problem of heat kernel function for $\Delta_f$ under some weight conditions.
\bt[Theorem \ref{deformedkernel}]\label{Thm1}Let $f_0$ be a non-degenerate quasi-homogeneous polynomial on $\C^n$ such that $q_M-q_m<\frac{1}{3}$, and let $f=f(z,u)=f_0(z)+\sum_{i=1}^su^i\phi_i(z)$ be one of the following deformations:
\begin{enumerate}
	\item the marginal deformation of $f_0$;
	\item the relevant deformation of $f_0$ with one more weight condition $\mbox{wt}(\phi_i)<1-2(q_M-q_m)$.
\end{enumerate}
Let $\Delta_f$ act on the Schwartz form space $\mathcal{S}\mathcal{A}_M^*(\C^n)$, with the norm $\|\cdot\|_0$. Then there exists a unique heat kernel function $p(z,w,t;u)\footnote{The function $p$ is a fuction of $t,z,w,u,\bar{z},\bar{w},\bar{u}$.}$ for the operator $\Delta_{f}$, which smoothly depends on the deformation parameters.

In particular, when $f_0$ is homogeneous, the result holds for its relevant or marginal deformation.
\et

Through $L^2$ extension, we consider the heat trace of $e^{-t\Delta_f}$, define the $i$-theta function $\zeta_f^i(s)$ associated to $\Delta_f$ and derive the vanishing theorem for $\zeta_f^1(s)$ as follows:
\bt[Theorem \ref{vanishf}] Under the weight conditions in Theorem \ref{Thm1},
$$\zeta_f^1(s)=0.$$
\et
The above result tells us that the vanishing result does not depend on the deformation parameters. It can also be viewed as the counterpart of the vanishing theorem of Ray-Singer analytic torsion, which also holds in the super-trace level
and does not depend on the complex structure of the underlying complex manifold. However, the proof here is more difficult, since the whole proof need start from the non-deformed case; there is also a significant difference between the CY B-model and LG B-model: there are left, right fermion operators (preserving the vacuum) in the CY case, which is given by counting the holomorphic and anti-holomorphic degree of the forms respectively, while the fermion operator in the LG case is just the sum of the holomorphic degree and the anti-holomorphic degree. The difference also makes the computation of the transgression formula more tedious; c.f. Section 5.
\vskip 0.1cm
Next, comparing with equations \eqref{K} and \eqref{BCOVF1}, we see that they look like the second torsion invariant $T^2(f)$. One can naturally hope that $T^2(f)$ can give us the genus 1 information of the LG B-model. Since we know that the genus 1 partition function in the CY 3-fold is determined by the holomorphic anomaly equation, we hope to derive some similar anomaly equation. First in the super-trace level, we have

\bt[Transgression Formula, Theorem \ref{Transgress}] Let $N$ be the operator acting on $\mathcal{A}^{*}(\C^n)$ such that $N\big|_{\mathcal{A}^{k}}=k\emph{Id}$, and $f$ as above. Then
\begin{equation*}
\begin{aligned}
\bar{\p}_{\bar{j}}\p_i\Tr(-1)^NN^2(e^{-t\Delta_f}-\Pi)~=~&-2t\frac{d}{dt}\Tr(-1)^N\phi_i\bar{\phi}_{\bar{j}}e^{-t\Delta_f}\\
                                                        ~&+2t\frac{d}{dt}\int_{0}^{t}\Tr(-1)^N\phi_i\Delta_fe^{-t'\Delta_f}\bar{\phi}_{\bar{j}}e^{-(t-t')\Delta_f}dt',
\end{aligned}
\end{equation*}
where $\Pi$ is the projection to harmonic forms.
\et

Combining it with the $tt^*$ geometry, we get the holomorphic anomaly equation for $T^2(f)$ as follows:
\bt[Anomaly Formula, Theorem \ref{af}] \label{af} Let $f_0(z)$ be a non-degenerate and homogeneous polynomial on $\C^n$, and let $f(z;u)=f_0(z)+\sum_{i=1}^{s}u^i\phi_i(z)$ be its marginal or relevant deformation.
Then 
\begin{equation}
\bar{\p}_{\bar{j}}\p_i\log T^2(f)=(-1)^n\tr C_i\bar{C}_{\bar{j}}-\left(\Tr(-1)^N\int_0^t\phi_ie^{-t'\Delta_f}\bar{\phi}_{\bj}e^{-(t-t')\Delta_f}dt'\right)_1.
\end{equation}
where $C_i=\Pi\circ\phi_i$, $\Bar{C}_{\bar{j}}=\Pi\circ\bar{\phi}_{\bar{j}}$, $()_1$ denotes the coefficient of $t$ in the power series expansion. In particular, when $f$ is the marginal deformation,
$$\left(\Tr(-1)^N\int_0^t\phi_ie^{-t'\Delta_f}\bar{\phi}_{\bj}e^{-(t-t')\Delta_f}dt'\right)_1=0.$$
\et

This paper is organized as follows. In Section 2, we recall some background and basic definitions about the LG B-model. In Section 3, we discuss the $tt^*$ geometry structure. In Section 4, we recall some heat kernel result in \cite{Fan2016Torsion} and derive the vanishing theorem for $\zeta_f^1(s)$. In Section 5, we focus on the computation of $\bap_{\bj}\p_i\Tr(-1)^NN^2(e^{-t\Delta_f}-\Pi)$. In Section 6, we give the definition of the singularity torsion invariant and prove the anomaly formula. The heat kernel analysis in the deformed case is given in Appendix A. \\

\section{Background and Basic Definitions}

\subsection{The Supersymmetry Algebra}

The Schr\"{o}dinger operator $\Delta_{f_0}$ in our study origins from the 1d version of the LG B model in physics. The input data of such a model is given by a pair $(X,f_0)$, where
\begin{itemize}
  \item $X$ is a non-compact complete K\"{a}hler manifold,
  \item $f_0$ is a holomorphic function on $X$.
\end{itemize}
$f_0$ plays the role of the superpotential of 2d supersymmetric quantum field theory. In what follows, we will focus on the following case
%$$\boxed{ X=\mathbb{C}^n, \text{ and } f_0 \text { is a nondegenerate and quasi-homogeneous polynomial on } \mathbb{C}^n.}$$
$$X=\mathbb{C}^n, \text{ and } f_0 \text { is a nondegenerate and quasi-homogeneous polynomial on } \mathbb{C}^n.$$
In \cite{Cecotti1991Topological}, Ceccoti and Vafa studied the fusion of 2d topological and anti-topological theory from $(X,f_0)$, and got a $tt^*$ geometry structure. Most of quantities in such a rich geometry structure can be computed by dimension reduction to 1 dimension, i.e., to Supersymmetric Quantum Mechanics. In the Schr\"{o}dinger representation, the Hilbert space is given by $L^2\mathcal{A}(X)$, i.e. the space of $L^2$ integrable forms on $X$, and the charge operators $Q_{+},Q_{-}, Q_{+}^{\dag},Q_{-}^{\dag}$ are represented by
\begin{flalign*}
&Q_{+}=\bar{\p}_{f_0}:=\bar{\p}+df_0\wedge,\quad Q_{-}=\p_{f_0}:=\p+d\bar{f_0}\wedge,\\
&Q_{+}^{\dag}=\bar{\p}_{f_0}^{\dag}=:-\ast \p_{-f_0}\ast, \quad Q_{-}^{\dag}=\p_{f_0}^{\dag}=:-\ast \bar{\p}_{-f_0}\ast,
\end{flalign*}
where we have fixed a certain constant hermitian metric on $X$, and $\ast$ is the Hodge star operator acting on $\Lambda^*(X)$ with respect to the metric. Then we have the supersymmetric algebra structure
\begin{equation}
\begin{aligned}
&\p_{f_0}^2=\p_{f_0}^{\dag 2}=\overline{\p}_{f_0}^2=\overline{\p}_{f_0}^{\dag 2}=0,\\
&\{\p_{f_0},\p_{f_0}^{\dag}\}=\Delta_{f_0},\quad \{\overline{\p}_{f_0},\overline{\p}_{f_0}^{\dag}\}=\Delta_{f_0},\\
&\{\overline{\p}_{f_0},\p_{f_0}\}=\{\overline{\p}_{f_0}^{\dag},\p_{f_0}^{\dag}\}=\{\p_{f_0},\overline{\p}_{f_0}^{\dag}\}=\{\overline{\p}_{f_0},\p_{f_0}^{\dag}\}=0.
\end{aligned}
\end{equation}
The fermionic operator $N$ acts on $\Lambda^{*}(X)$, that is, for any $\alpha\in\Lambda^{k}(X)$, $N(\alpha)=k\alpha$. Then
\begin{equation}\label{N0}
[N,\overline{\p}_{f_0}]=\overline{\p}_{f_0},~[N,\p_{f_0}]=\p_{f_0},~[N,\overline{\p}_{f_0}^{\dag}]=-\overline{\p}_{f_0}^{\dag},~[N,\p_{f_0}^{\dag}]=-\p_{f_0}^{\dag},~[N,\Delta_{f_0}]=0.
\end{equation}

\noindent\textbf{Remark:} $\{,\}$ here is the anti-commutative bracket. Physicists always denote by $F$ the fermionic operator, c.f. the introduction. Here we adopt the mathematical notation $N$, i.e. the number operator.

\subsection{Basic definitions}
In this subsection, we give some basic definitions about the quasi-homogeneous polynomials and their deformations.
\bd $f_0\in\C[z_1,\ldots,z_n]$ is called a quasi-homogeneous polynomial, if there exist $q_1,\ldots,q_n\in\mathbb{Q}^+$, such that for any $\lambda\in\C^*$,
$$f_0(\lambda^{q_1}z_1,\ldots, \lambda^{q_n}z_n)=\lambda f_0(z_1,...,z_n).$$
Each $q_i$ is called the weight of $z_i$.
\ed

\bd Let $f_0\in\C[z_1,\ldots,z_n]$ be a quasi-homogeneous polynomial, it is called non-degenerate if
\begin{enumerate}
  \item $f_0$ contains no monomial of the form $z_iz_j$ for $i\neq j$,
  \item $f_0$ has only an isolated singularity at the origin.
\end{enumerate}
\ed

In the LG B model, the state space can be described by the Jacobi ring $\mathcal{J}(f_0)$ of $f_0$:
$$\mathcal{J}(f_0):=\C[z_1,\ldots,z_n]/\langle\frac{\p f_0}{\p z_1},\ldots,\frac{\p f_0}{\p z_n}\rangle.$$
In fact, in this case, $\mathcal{J}(f_0)$ is finite dimensional and denote $\mu=\mu_{f_0}=\dim \mathcal{J}(f_0)$. Now we take a monomial basis $\phi_{1},...,\phi_{\mu}$ of $\mathcal{J}(f_0)$, choose arbitrary $s$ elements from $\{\phi_{j}\}_{j=1}^{\mu}$, for example, $\phi_{1},\ldots,\phi_{s}$, and define
$$f(z;u)=f_{0}+\sum_{j=1}^{s}u_{j}\phi_{j}.$$
Then $f$ is called a deformation of $f_0$.

Since $f_0$ has weight 1, we can think of $f$ as a quasi-homogeneous polynomial of $(z,u)$, that is, $\mbox{wt}(u_j):=1-\mbox{wt}(\phi_j)$.
\bd
The deformation parameter $u_{j}$ is called:
\begin{enumerate}
  \item relevant, if the weight of $u_j$ is positive;
  \item marginal, if the weight of $u_j$ is zero;
  \item irrelevant, if the weight of $u_j$ is negative.
\end{enumerate}
\ed

We will focus on the relevant deformation and marginal deformation, since the two cases do not change the singular property of $f_0$ at $\infty$.

\bex\label{An} (1) For the $A_n$ singularity $f_{0}=\frac{1}{n+1}z^{n+1}$, one can deform $f_0$ to be
$$f=f_{0}+u_{n-1}z^{n-1}+u_{n-2}z^{n-2}+\cdots+u_{1}z+u_{0}.$$
It is a relevant deformation.
\vskip 0.1cm
(2) Let $f_{0}=z_{1}^{n}+\cdots+z_{n}^{n}$. If we view $[z_1,\ldots,z_n]$ as the homogeneous coordinates of $\mathbb{CP}^{n-1}$, then $f_{0}=0$ gives a complex hypersurface in $\mathbb{CP}^{n-1}$. There is a common deformation
$$f=f_{0}+uz_{1}\cdots z_{n},$$
which is a marginal deformation. Such a deformation has a global $\C^{*}$ action, and hence the marginal deformation induces the complex deformation of the complex hypersurface.
\eex
More definitions and examples can be found in Section 3 of \cite{Fan2011Schr}.\\

\section{$tt^*$ geometry}

In this section, first we consider the relation between the cohomology spaces attached to $\bar{\p}_{f_0}$ and the space of harmonic forms attached to $\Delta_{f_0}$. Then, applying this relation to the deformed case, we introduce the genus 0 theory of the LG B-model, which is governed by the $tt^*$ geometric structure on the space of the deformation parameters.

\subsection{Hodge decomposition}

We recall the following results in \cite{Fan2011Schr}.

For simplicity, let us introduce some notations:
\begin{itemize}
  \item $\H_0^*:=\ker(\Delta_{f_0}:L^2\mathcal{A}^*(X)\rightarrow L^2\mathcal{A}^*(X))$ is the space of harmonic forms on $X$.
  \item $H^{*}_{\overline{\p}_{f_{0}}}(X)$ is the cohomology of the smooth complex $(\mathcal{A}^*(X), \overline{\p}_{f_{0}})$.
  \item $H^{*}_{(2),\overline{\p}_{f_{0}}}(X)$ is the cohomology of the $L^2$-complex $(L^2\mathcal{A}^*(X), \overline{\p}_{f_{0}})\footnote{$\bar{\p}_{f_0}$ is defined in the sense of distribution. One can also consider the smooth $L^2$ complex, Fan\cite{Fan2011Schr} proved that the two cohomology groups are isomorphic via some regularization.}$.
\end{itemize}

\noindent In this subsection, we state the relations among the three spaces.

First, it is easy to compute the cohomology group $H^{*}_{\overline{\p}_{f_{0}}}(X)$ via the spectral sequence, which is given by
$$H^{*}_{\overline{\p}_{f_{0}}}(X)=H^*(\mathcal{A}^*(X), \overline{\p}_{f_{0}})=\O^{n}(X)/df_{0}\wedge\O^{n-1}(X)\cong\mathcal{J}(f_0).$$

To get the information of $\H_0^*$ and $H^{*}_{(2),\overline{\p}_{f_{0}}}(X)$, we first have the following analytical result for $\Delta_{f_0}$.
\bt[Theorem 2.40 in \cite{Fan2011Schr}] \label{spec}
Suppose that $f_{0}$ is a non-degenerate and quasi-homogeneous polynomial on $\mathbb{C}^n$, then the form Laplacian $\Delta_{f_{0}}$ on $L^{2}\mathcal{A}^{*}(X)$ has purely discrete spectrum and all the eigenforms form a complete basis of the Hilbert space $L^2\mathcal{A}^*(X)$.
\et

Furthermore, we have the Hodge decomposition for $L^{2}\mathcal{A}^{*}(X)$:
\bt[Theorem 2.52 in \cite{Fan2011Schr}]\label{Hodge}
There are orthogonal decompositions
$$ L^{2}\mathcal{A}^{*}(X)=\H_{0}^{*}\oplus \emph{im}(\overline{\p}_{f_{0}})\oplus \emph{im}(\overline{\p}_{f_{0}}^{\dag}), $$
$$\ker\overline{\p}_{f_{0}}=\H_{0}^{*}\oplus \emph{im}(\overline{\p}_{f_{0}}).$$
In particular, we have the isomorphism
$$H^{*}_{(2),\overline{\p}_{f_{0}}}(X)\cong \H_{0}^{*}.$$
\et

At last, the following theorem reveals the relation between $\H^{*}$ and $H^{*}_{\overline{\p}_{f_{0}}}(X)$.
\bt[Theorem 2.63 in \cite{Fan2011Schr}] \label{dimH}
Let $(X, f_0)$ be as before. Then
\begin{equation}
\label{eq:abs}
\dim \H_{0}^{k}=
\begin{cases}
0& \text{if $k \neq n$}\\
\mu & \text{if $k=n$}.
\end{cases}
\end{equation}
And there is an explicit isomorphism:
$$i:\H_{0}^{n}\longrightarrow \O^{n}(X)/df_{0}\wedge\O^{n-1}(X)\cong\mathcal{J}(f_0).$$
\et

\noindent\textbf{Remark:} To construct the isomorphism $i$, it is more convenient to introduce the cohomology $H^{*}_{c,\overline{\p}_{f_{0}}}(X)$ of the compact-supported smooth complex $(\mathcal{A}_c^*(X), \overline{\p}_{f_{0}})$. One can show that there is an explicit quasi-isomorphism between $(\mathcal{A}^*(X), \overline{\p}_{f_{0}})$ and $(\mathcal{A}_c^*(X), \overline{\p}_{f_{0}})$, see for example in \cite{Li2013Primitive}. Then, one can construct the element in $H^{*}_{(2),\overline{\p}_{f_{0}}}(X)$ from $H^{*}_{c,\overline{\p}_{f_{0}}}(X)$. Finally, there exists a corresponding harmonic form via Theorem \ref{Hodge}, from which we get the map $i$.
\vskip 0.1cm
In the next subsection, we consider those cohomologies in the deformed case.

\subsection{$tt^*$ equations: the genus 0 anomaly equations}

Let $\phi_{1},\phi_{2},...\phi_{\mu}$ be a monomial basis of the Jacobi ring $\mathcal{J}(f_0)=\C[z_1,z_2,...,z_n]/(f_0'(z))$. Let
$$f(u,z)=f_{0}(z)+\sum_{i=1}^{s}u^i\phi_{i}(z)$$
be the relevant or marginal deformation.\footnote{We use the upper indices for the parameters in the sequel.}
Denote by $M$ the space of parameters $(u^1,\ldots,u^{s})$, which is a small neighborhood of the origin in $\C^{s}$. We therefore have a family of supersymmetric algebra operators $\bar{\p}_{\b f},\p_{\b f}, \bar{\p}^{\dag}_{\b f},\p_{\b f}^{\dag},\Delta_{\b f}$ parameterized by $(\b,u)\in \mathbb{C}^{*}\times M$. When $|u|$ is small enough, the dimension of the Jacobi ring of $f$ remains the same, and the results in the last subsection hold for $\Delta_{\b f}$.
\vskip 0.1cm
We now consider the $tt^*$ geometry structure on the Hodge bundle over $M$.

We first have the trivial complex Hilbert bundle $\Lambda_{M}^{*}(X):=L^{2}\Lambda^{*}(X)\times \mathbb{C}^{*}\times M\rightarrow \mathbb{C}^{*}\times M$. For simplicity, denote by $\mathcal{A}_{M}^*(X)$ the section space of $\Lambda_{M}^{*}(X)$.  There is a natural Hermitian metric
\begin{flalign*}
g: \mathcal{A}_{M}^*(X)\times \mathcal{A}_{M}^*(X)&\longrightarrow C^{\infty}(\C^*\times M)\\
(\a_1,\a_2)&\longmapsto\int_{X}\a_1\wedge\ast\overline{\a_2}.
\end{flalign*}

Note that there is a canonical real structure on the Hilbert bundle $\Lambda_{M}^{*}(X)$ which is given by the complex conjugate, and we denote it by $\tau_{R}$. By Theorem \ref{spec}, for each $(\b,u)\in\C^*\times M$, we can choose the eigenforms $\{\a_{a}(\b,u)\}_{a=1}^{\infty}$ of $\Delta_{\b f(u)}$ to form a basis of $L^{2}\mathcal{A}^{*}(X)$. Then $\{\a_a=\a_{a}(\b,u)\}_{a=1}^{\infty}$ forms a frame of $\Lambda_{M}^*(X)$ and $\{\overline{\a_{a}}\}_{a=1}^{\infty}$ also forms a frame of $\Lambda_{M}^*(X)$.

Let $H_{M}^{n}\subset\Lambda_{M}^{*}(X)$ be the Hodge bundle over $\C^{*}\times M$, and the fiber at $(\b,u)\in\C^{*}\times M$ is given by the space of all harmonic $n$ forms of $\Delta_{\b f(u)}$. The space of section is denoted by $\H^n$.

In the sequel, we restrict the metric $g$ to the Hodge bundle and still denote it by $g$, which is called the $tt^*$ metric.

\noindent\textbf{Remark:} As we mentioned in the last subsection, the cohomology group $H^{*}_{(2),\overline{\p}_{\b f}}(X)$ is isomorphic to $\H^n$ as vector spaces. Naturally, one can ask whether the metric $g$ could descend to the cohomology elements. Unfortunately, there exists an ambiguity: i.e. there exist $\a_1\neq\a_2$ belonging to the same cohomology class, and a $\bar{\p}_{f}$-closed form $\a_3$, such that $g(\a_1,\a_3)\neq g(\a_2,\a_3)$. To fix the ambiguity, one need consider the harmonic forms instead. For this reason, it becomes much more difficult to compute the $tt^*$ metric.

\vskip 0.1cm

Assume that $\{\a_a\}_{a=1}^{\mu}$ form a basis of $\H^n$, then from the Hodge decomposition, for any $\a\in\mathcal{A}_{M}^n(X)$, there exist unique $\xi\in\H^n, \gamma_1\in\mbox{im}(\bar{\p}_{\b f}), \gamma_2\in\mbox{im}(\bar{\p}_{\b f}^{\dag})$, such that
$$\a=\xi+\gamma_1+\gamma_2,$$
and $g(\a, \a_a)=g(\xi, \a_a)$ for any $a=1,\ldots,\mu$.

By Theorems \ref{spec} and \ref{Hodge}, we know that $\Delta_{\b f}$ is invertible on the subspace $\mbox{im}(\bar{\p}_{\b f})\oplus\mbox{im}(\bar{\p}_{\b f}^{\dag})$. Denote the inverse operator by $G$. Let $\Pi: \mathcal{A}_{M}^*(X)\rightarrow \H^n$ be the harmonic projection. Then the Hodge decomposition can be written in the operator form :
$$\mbox{Id} = \Pi+G\Delta_{\b f}=\Pi+\Delta_{\b f}G.$$
It is easy to check that $G$ commutes with $\bar{\p}_{\b f},\p_{\b f}, \bar{\p}^{\dag}_{\b f},\p_{\b f}^{\dag},\Delta_{\b f}$.

\vskip 0.1cm
Now we can define some important operators on the Hodge bundle:\\

\noindent{1. The connections $D$, $\bar{D}$}
\vskip 0.1cm
Notice that the Hodge bundle is embedded into the Hilbert bundle, so we can define $D$ in a natural way:
$$D_i=\Pi\circ \p_i,\quad \bar{D}_{\bar{i}} =\Pi\circ\bar{\p}_{\bar{i}}\quad i=1,...,s.$$
Or equivalently we can define their components:
$$(D_{i})_{a\bar{b}}=g(\p_{i} \a_{a},\a_{b}), \quad (\bar{D}_{\bar{i}})_{a\bar{b}}=g(\bap_{\bar{i}}\a_{a},\a_{b}),$$
then $D_{i}\a_{a}=(D_{i})_{a\bar{b}}g^{\bar{b}c}\a_{c}, \bar{D}_{\bar{i}}\a_{a}=(\bar{D}_{\bar{i}})_{a\bar{b}}g^{\bar{b}c}\a_{c}.$

Using the relations $[\p_i,\bar{\p}_{\b f}]=\p_{\b f}(\p_i\b f), \quad [\p_i, \bar{\p}_{\b f}^{\dag}]=0$, we have
\begin{flalign*}
D_i\a&=\Pi\p_i\a=\p_i\a-G\Delta_{\b f}\p_i\a\\
     &=\p_i\a-G[\Delta_{\b f},\p_i]\a\quad(\Delta_{\b f}\a=0)\\
     &=\p_i\a+G\bar{\p}_{\b f}^{\dag}\p_{\b f}(\p_i\b f)\a\quad\quad (\bar{\p}_{\b f}\a=\bar{\p}_{\b f}^{\dag}\a=0)\\
     &=\p_i\a+G\bar{\p}_{\b f}^{\dag}\p_{\b f}[(\p_i\b f)\a]\quad\quad (\p_{\b f}\a=0)\\
     &=\p_i\a-\b\p_{\b f}\bar{\p}_{\b f}^{\dag}G(\p_if)\a.
\end{flalign*}
Similarly, $\bar{D}_{\bar{i}}\a=\p_{\bar{i}}\a_{a}-\bar{\b}\bar{\p}_{\b f}\p^{\dag}_{\b f}G(\overline{\p_{i}f})\a$.
\vskip 0.1cm

\noindent{2. The operators $C_{i}, \bar{C}_{\bar{i}}$}
\vskip 0.1cm
Define $C_{i}=\Pi\circ\p_{i}f$, $\bar{C}_{\bar{i}}=\Pi\circ\overline{\p_if}$,
or equivalently define their components:
$$(C_{i})_{a\bar{b}}=g(\p_{i}f \a_{a},\a_{b}), \quad(\bar{C}_{\bar{i}})_{a\bar{b}}=g(\overline{\p_{i}{f}}\a_{a},\a_{b}).$$
Similarly, we have the formula
$$C_{i}\a_{a}=(\p_{i}f\a_{a})-\bar{\p}_{\b f}\bar{\p}^{\dag}_{\b f}G(\p_{i}f)\a_{a}, \quad \bar{C}_{\bar{i}}\a_{a}=(\overline{\p_{i}f}\a_{a})-\p_{\b f}\p^{\dag}_{\b f}G(\overline{\p_{i}f})\a_{a}. $$

\bpr
The operators $D,\bar{D},C,\bar{C}$ satisfy the following relations:
\begin{enumerate}
  \item $[C_{i},C_{j}]=[\bar{C}_{\bar{i}}, \bar{C}_{\bar{j}}]=0$;
  \item $[D_{i},\bar{C}_{\bar{j}}]=[\bar{D}_{\bar{i}},C_{j}]=0$;
  \item $[D_{i},C_{j}]=[D_{j},C_{i}], \quad [\bar{D}_{\bar{i}}, \bar{C}_{\bar{j}}]=[\bar{D}_{\bar{j}},\bar{C}_{\bar{i}}]$;
  \item $[D_i,D_j]=[\bar{D}_{\bar{i}},\bar{D}_{\bar{j}}]=0, \quad [D_{i},\bar{D}_{\bar{j}}]=-|\b|^{2}[C_{i},\bar{C}_{\bar{j}}].$
\end{enumerate}
\epr
\bp For simplicity, denote by $\bar{\p}_{\b f}\gamma$ (or $\p_{\b f}\gamma$) the element that we need in im$\bar{\p}_{\b f}$ (or im$\p_{\b f}$), because we always need project to the harmonic part. For any $\a\in \mathcal{H}^n$, to show
\vskip 0.1cm
(1) $[C_i, C_j]=0:$
$$\p_{i}f\p_{j}f\a=\p_{i}f(C_{j}\a+\bar{\p}_{\b f}\gamma)=C_{i}C_{j}\a+\bar{\p}_{\b f}\gamma.$$
Where we used the relation $[\p_{i}f,\bar{\p}_{\b f}]=0$. Similarly,
$$\p_{j}f\p_{i}f\a=C_{j}C_{i}\a+\bar{\p}_{\b f}\gamma.$$
So $C_{i}C_{j}\a=C_{j}C_{i}\a.$ The proof of the rest equalities is the same.
\vskip 0.1cm
(2) $[\bar{D}_{\bar{i}}, C_j]=0:$
\begin{equation*}
\begin{aligned}
\bap_{\bar{i}}C_{j}\a &=\bap_{\bar{i}}(\p_{j}f\a+\bar{\p}_{\b f}\gamma)\\
                    &=\p_{j}f\bap_{\bar{i}}\a+\bar{\p}_{\b f}\gamma=\p_{j}f(\bar{D}_{\bar{i}}\a+\bar{\p}_{\b f}\gamma)+\bar{\p}_{\b f}\gamma\\
                    &=\p_{j}f\bar{D}_{\bar{i}}\a+\bar{\p}_{\b f}\gamma=C_{j}\bar{D}_{\bar{i}}\a+\bar{\p}_{\b f}\gamma.
\end{aligned}
\end{equation*}
In the second equality, we use $[\p_{\bar{i}},\bar{\p}_{\b f}]=0$ and in the third equality we use $[\p_{j}f,\bar{\p}_{\b f}]=0$.
So $[D_{i},\bar{C}_{\bar{j}}]=0$. Analogously, $[D_i,\bar{C}_{\bj}]=0$.
\vskip 0.1cm
(3) $[D_i, C_j]=[D_j, C_i]:$
\begin{equation*}
\begin{aligned}
\p_{i}C_{j}\a &=\p_{i}(\p_{j}f\a-\bar{\p}_{\b f}\bar{\p}^{\dag}_{\b f}G(\p_{j}f)\a)\\
              &=(\p_{i}\p_{j}f)\a+\p_{j}f\p_{i}\a+\bar{\p}_{\b f}\gamma-\b\p_{i}(\p f\wedge)\bar{\p}^{\dag}_{\b f}G(\p_{j}f)\a\\
              &=(\p_{i}\p_{j}f)\a+\p_{j}f(D_{i}\a+\b\p_{\b f}\bar{\p}^{\dag}_{\b f}G(\p_{i}f)\a)+\bar{\p}_{\b f}\gamma-\b\p_{i}(\p f\wedge)\bar{\p}^{\dag}_{\b f}G(\p_{j}f)\a\\
              &=(\p_{i}\p_{j}f)\a+\p_{j}f D_{i}\a-\b\p_j(\p f\wedge)\bar{\p}^{\dag}_{\b f}G(\p_{i}f)\a+\p_{\b f}\gamma+\bar{\p}_{\b f}\gamma-\b\p_{i}(\p f\wedge)\bar{\p}^{\dag}_{\b f}G(\p_{j}f)\a.
\end{aligned}
\end{equation*}
So $$D_{i}C_{j}\a=C_{j}D_{i}\a+\Pi\left((\p_{i}\p_{j}f)\a-\b\p_j(\p f\wedge)\bar{\p}^{\dag}_{\b f}G(\p_{i}f)\a-\b\p_{i}(\p f\wedge)\bar{\p}^{\dag}_{\b f}G(\p_{j}f)\a\right).$$
And in the second term, it is obviously that $i$ and $j$ are symmetry. So $[D_{i},C_{j}]\a=[D_{j},C_{i}]\a$. Analogously, $[\bar{D}_{\bar{i}}, \bar{C}_{\bar{j}}]=[\bar{D}_{\bar{j}},\bar{C}_{\bar{i}}]$.

\vskip 0.1cm
(4) $[D_i, D_j]=0:$
\begin{equation*}
\begin{aligned}
\p_{i}D_j\a &=\p_{i}(\p_j\a-\b\p_{\b f}\bar{\p}_{\b f}^{\dagger}G(\p_{j}f)\a)\\
            &=\p_i\p_j\a-\p_{\b f}\gamma. \quad ([\p_i,\p_{\b f}]=0.)
\end{aligned}
\end{equation*}
So $[D_i,D_j]=0$. Analogously, $[\bar{D}_{\bar{i}}, \bar{D}_{\bar{j}}]=0$.
\vskip 0.1cm
The last and also the most important formula is about the curvature $[D_i, \bar{D}_{\bar{j}}]$:
\begin{flalign*}
\p_{i}\bar{D}_{\bar{j}}\a &=\p_{i}(\bap_{\bar{j}}\a-\bar{\b}\bar{\p}_{\b f}\p_{\b f}^{\dagger}G\overline{\p_{j}f}\a)\\
                    &=\p_{i}\bap_{\bar{j}}\a-\bar{\b}\p_{i}\bar{\p}_{\b f}\p_{\b f}^{\dagger}G\overline{\p_{j}f}\a\\
                    &=\p_{i}\bap_{\bar{j}}\a-\bar{\p}_{\b f}\gamma-|\b|^{2}(\p_{i}(\p f\wedge))\p_{f}^{\dagger}G\overline{\p_{j}f}\a.
\end{flalign*}
In the second row, we use $[\p_{i},\bar{\p}_{\b f}]=\b\p_{i}(\p f)$. So
$$D_{i}\bar{D}_{\bar{j}}\a=\Pi\left(\p_{i}\bap_{\bar{j}}\a-|\b|^{2}(\p_{i}(\p f\wedge))\p_{\b f}^{\dagger}G\overline{\p_{j}f}\a\right).$$
Similarly,
$$\bar{D}_{\bar{j}}D_{i}\a=\Pi\left(\bap_{\bar{j}}\p_{i}\a-|\b|^{2}(\bap_{\bar{j}}(\overline{\p f}\wedge))\bar{\p}_{\b f}^{\dagger}G(\p_{i}f)\a\right).$$

On the other hand,
\begin{equation*}
\begin{aligned}
\p_{i}f \bar{C}_{\bar{j}}\a&=\p_{i}f(\overline{\p_{j}f}\a-\p_{\b f}\p_{\b f}^{\dagger}G\overline{\p_{j}f}\a)\\
                     &=\p_{i}f\overline{\p_{j}f}\a-\p_{\b f}\gamma+(\p_i(\p f\wedge))\p_{\b f}^{\dagger}G\overline{\p_{j}f}\a.
\end{aligned}
\end{equation*}
In the second row, we use $[\p_{i}f,\p_{\b f}]=-\p(\p_{i}f)\wedge.$
So, $$C_{i}\bar{C}_{\bar{j}}\a=\Pi\left(\p_{i}f\overline{\p_{j}f}\a+(\p_i(\p f\wedge))\p_{\b f}^{\dagger}G\overline{\p_{j}f}\a\right).$$
Similarly,
$$\bar{C}_{\bar{j}}C_{i}\a=\Pi\left(\overline{\p_{j}f}\p_{i}f\a+(\bar{\p}_{\bar{j}}(\overline{\p f})\wedge)\bar{\p}_{\b f}^{\dagger}G(\p_{i}f)\a\right).$$
Now it is obvious that $[D_{i},\bar{D}_{\bar{j}}]\a=-|\b|^{2}[C_{i},\bar{C}_{\bar{j}}]\a.$
\ep

\noindent\textbf{Remarks:} (1) The equations in this proposition are called the $tt^*$ equations, which is equivalent to that the connection $\nabla=D+\bar{D}+\b C_{i}du^i+\bar{\b}\bar{C}_{\bar{i}}du^{\bar{i}}$ on the Hodge bundle is flat. The $tt^*$ equations are a generalization of the special geometry relation on the Calabi-Yau 3-fold which is famous as the genus 0 anomaly equations in \cite{Bershadsky1994Kodaira}.
\vskip 0.1cm
(2) It is possible to choose a holomorphic gauge such that the connection matrices $A_i$, $\bar{A}_{\bar{i}}$ of $D_i$, $\bar{D}_{\bar{i}}$ respectively reads
$$\bar{A}_{\bar{i}}=0,\quad A_i=-g\p_ig^{-1}.$$
So $D$ is the Chern connection of the Hermitian metric $g$.
\vskip 0.1cm
(3) One can also define more operators $\mathcal{U}$, $\bar{\mathcal{U}}$, $D_{\b}$, $\bar{D}_{\bar{\b}}$ to be
$$\mathcal{U}=\Pi\circ f, \quad \bar{\mathcal{U}}=\Pi\circ\bar{f},\quad D_{\b}=\Pi\circ\p_{\b}, \quad \bar{D}_{\bar{\b}}=\Pi\circ\bap_{\bar{\b}}.$$
Then one can prove that the extended connection
$$\widetilde{\nabla}:=\nabla+D_{\b}d\b+\bar{D}_{\bar{\b}}d\bar{\b}+\mathcal{U}d\b+\bar{\mathcal{U}}d\bar{\b}$$
is flat. The complete discussion could be found in \cite{Fan2011Schr}.

\bd[$tt^*$ geometry] A $tt^*$ geometry structure $(K\rightarrow M, \kappa, g, D, C, \bar{C})$ consists of the following data
\begin{itemize}
  \item $K\rightarrow M$ is a smooth vector bundle,
  \item a complex anti-linear involution $\kappa: K\rightarrow K$, i.e. $\kappa^2=\Id$, $\kappa(\l\a)=\bar{\l}\kappa(\a)$, $\forall \l\in\C$,
  \item a Hermitian metric $g(u,v)$,
  \item a one parametric family of flat connections $\nabla^z=D+\bar{D}+\frac{1}{z}C+z\bar{C}$, where $D+\bar{D}$ is the Chern connection of $g$, $C$, $\bar{C}$ are the $C^{\infty}(M)$-linear map
      $$C:C^{\infty}(K)\rightarrow C^{\infty}(K)\otimes\m{A}_M^{(1,0)},\quad \bar{C}:C^{\infty}(K)\rightarrow C^{\infty}(K)\otimes\m{A}_M^{(0,1)}$$
\end{itemize}
satisfying
\begin{enumerate}
  \item $g$ is real with respect to $\kappa: g(\kappa(u),\kappa(v))=\overline{g(u,v)}$,
  \item $D+\bar{D}$ respects the Hermitian metric $g$,
  \item $(D+\bar{D})(\kappa)=0$, $\bar{C}=\kappa\circ C\circ\kappa$,
  \item $\bar{C}$ is the adjoint of $C$ with respect to $g$, i.e. $g(C_Xu,V)=g(u,\bar{C}_{\bar{X}}v)$, $X\in\m{T}_M$.
\end{enumerate}
Then we say such structure $(K\rightarrow M, \kappa, g, D, C, \bar{C})$ is a $tt^*$ geometry structure.
\ed

One can show that $(\H^n\rightarrow M, \tau_R, g, D, \bar{D}, C=C_idu^i, \bar{C}=\bar{C}_{\bar{i}}du^{\bar{i}})$ admits a $tt^*$ geometry structure. If we impose the condition that $\dim(M)=\mu$, more precisely, when the central charge $\widehat{c}=\sum_{i=1}^n(1-2q_i)<1$, $M$ can be the universal deformation space, the $tt^*$ geometry structure captures the genus 0 information of the LG B-model. It is parallel to the Frobenius manifold structure in the CY or LG A-model. In what follows, we will mainly discuss the ``genus 1" anomaly equations for the LG B-model through the heat kernel analysis.\\

\section{Heat Kernel Analysis and Zeta functions}

In this section, we first recall some main results in \cite{Fan2016Torsion}, which consists of the existence of heat kernel function for $\Delta_{f_0}$, the heat trace problem and the vanishing theorem attached to $\Delta_{f_0}$. In the following discussion, we still focus on the relevant deformation or marginal deformation. In the last subsection, we will extend the vanishing result to the deformed case. Note that we set $\beta=1$ in the sequel, but the result still holds if we regard $\b\in\C^*$ as a parameter.

\subsection{Heat kernel for $\Delta_{f_0}$ and $\Delta_{f}$}

Let $f_0$ be a non-degenerate and quasi-homogeneous polynomial on $\C^n$, and let $h$ be a fixed constant Hermitian metric on $\C^n$. Then $\Delta_{f_0}$ has the local expression

$$\Delta_{f_0}=-\Delta_{\bar{\p}}+L_{f_0}+|\nabla f_0|^{2},$$
where
$$L_{f_0}=-(h^{\bar{\mu}\nu}\p_{\nu}{f_0}_{l}\iota_{\p_{\bar{\mu}}}dz^{l}\wedge+\overline{h^{\bar{\mu}\nu}\p_{\nu}{f_0}_{l}\iota_{\p_{\bar{\mu}}}dz^{l}\wedge}).$$

Denote by $q_M$ and $q_m$ the maximal and minimal weight of variables in $f_0$ respectively. In \cite{Fan2016Torsion}, Fan and Fang constructed the heat kernel function of $e^{-t\Delta_{f_0}}$ by approximation and iteration method for $q_M-q_m<\frac{1}{3}$. Most of singularities which interest us satisfy this weight condition, for example, all homogeneous polynomials, ADE singularities, the unimodal singularities. Of course, it is easy to find a non-degenerate and quasi-homogeneous polynomial which does not meet the requirement, e.g. $x^2+xy^3, x^2+xy^2+yz^4$.
\vskip 0.1cm
The existence problem of the heat kernel function for $\Delta_{f_0}$ is solved by the following.

\bt [Theorem 5.13 in \cite{Fan2016Torsion}] \label{kernel}
Let $f_0$ be a non-degenerate quasi-homogeneous polynomial on $\mathbb{C}^{n}$ satisfying $q_{M}-q_{m}<\frac{1}{3}$ and let $l, K\in \mathbb{N}$, $\delta=\frac{1-3(q_M-q_m)}{1-q_M}$ satisfy $\frac{K\delta}{3}-\frac{5-9q_m}{3(1-q_M)}-n-l_{0}-\frac{l_{0}q_M}{2-2q_M}>0$. Fix $T>0$, then for any $t\in (0,T]$, the series
$$p(z,w,t)=\sum_{i=0}^{\infty}(-1)^{i}p_{K}^{i}(z,w,t)\footnote{The kernel function $p$ is a function of $z,w,\bar{z},\bar{w},t$, for simplity, we omit the $\bar{z},\bar{w}$.}$$
converges for any $(z,w)\in \mathbb{C}^{n}\times\mathbb{C}^{n}$, and moreover has up to $l_0$-order $z$-derivatives and at least 1-order $t$-derivatives. $p(z,w,t)$ is the unique heat kernel function for the operator $\Delta_{f_0}$.
\et
\bp See \cite{Fan2016Torsion} or Appendix A\footnote{We rewrite Fan-Fang's proof in a simple way, which will be necessary for the following analysis, such as $L^2$ extension, asymptotic analysis.} in this paper.
\ep

\noindent\textbf{Remark:} The existence of the heat kernel function for the following two operators $L$ are well-known:
\begin{enumerate}
  \item The base space is $\mathbb{R}^n$, $L=\sum_{i=1}^n\p_{x_i}^2+V(x)$, where $V(x)$ has compact support;
  \item The base space is a compact smooth manifold, and $L$ is a second order elliptic operator.
\end{enumerate}
The two existence problems can be solved via the perturbative method. But the case here is different: the base space is non-compact, and the potential function $V(z,\bar{z})$ is bounded from below, $V\rightarrow+\infty$ as $|z|\rightarrow+\infty$. So we can not use the usual perturbative method. In order to ensure the convergence problem, we need do the resummation to get a control factor, the details can be found in Appendix A.\\

If we regard $u\in M$ just as parameters, then it is obvious that the same result holds for marginal deformation $f$. Actually, the existence problem of heat kernel function for the deformed case can be solved under some mild conditions as follows. Since the proof is a little more tedious, we put the details in Appendix A.

\bt\label{deformedkernel} Let $f_0$ be a non-degenerate quasi-homogeneous polynomial on $\C^n$ such that $q_M-q_m<\frac{1}{3}$, and let $f=f(z,u)=f_0+\sum_{i=1}^su^i\phi_i$ be one of the following deformations:
\begin{enumerate}
  \item the marginal deformation of $f_0$;
  \item the relevant deformation of $f_0$ with one more weight condition $\mbox{wt}(\phi_i)<1-2(q_M-q_m)$.
\end{enumerate}
Let $\Delta_f$ act on the Schwartz form space $\mathcal{S}\mathcal{A}_M^*(\C^n)$, with the norm $\|\cdot\|_0$. Then there exists a unique heat kernel function $p(z,w,t;u)\footnote{Similarly, the function $p$ also depends on $\bar{u}$.}$ for the operator $\Delta_{f}$, which smoothly depends on the deformation parameters.

In particular, when $f_0$ is homogeneous, the result holds for its relevant or marginal deformation.
\et
\bp The idea of the proof will be given in Appendix A, after the proof of Theorem \ref{kernel}.
\ep

For convenience, in the sequel, we use the notation $(\star)$ to denote the weight conditions for $f_0$ and $f$ in Theorem \ref{deformedkernel}.

Fan and Fang proposed that the condition $q_M-q_m<\frac{1}{3}$ is not necessary for the existence of heat kernel. For example, the heat kernel associated to $x^2+y^6$ exists, since the singularity is splittable. As we show in Appendix A, using the construction of parametrix in \cite{Fan2016Torsion} and some dimension argument, there is another sufficient condition available for the existence problem. That is, if the higher derivative (bigger than 1) of $f_0$ can be controlled by $C(|\nabla f_0|+1)$, where $C$ is some positive constant depending on $n$, then we can construct the heat kernel function for $\Delta_{f_0}$, even for the relevant or marginal deformed case $\Delta_f$. Through a simple calculation, we can see that $x^2+xy^3$ meets the condition. But in general, it is difficult to check this condition.

\subsection{Heat trace and zeta function}

\subsubsection{$L^2$ extension and the trace of the heat kernel}

Note that in the construction of $p(z,w,t;u)$ under the weight condition $(\star)$, following from the proof of Proposition \ref{estimater}, we can obtain
\begin{equation}\label{pt}
|p(z,w,t;u)|\leq H(t;u) \quad \text{ and } \quad \int_{\C^n}|p(z,w,t;u)|dw\wedge d\bar{w}<G(t;u).
\end{equation}
Then it is direct to get the following results from the inequalities \eqref{pt} and the Schwartz's inequality.
\bpr (1) For any $\psi(z;u)\in L^2\mathcal{A}_M^*(\C^n)$, we have $\int_{\C^n}p(z,w,t;u)\psi(w;u)\emph{dvol}_h(w)\in L^2\mathcal{A}_M^*(\C^n)$. In other words, $e^{-t\Delta_f}$ admits an $L^2$ extension with kernel function $p(z,w,t;u)$.
\vskip 0.1cm
(2) $p(z,w,t;u)$ is $L^2$-integrable. Thus $e^{-t\Delta_f}$ is a Hilbert-Schmidt operator for any $\re(t)>0$. Furthermore, $e^{-t\Delta_f}$ is trace-class and the trace is given by
$$\Tr(e^{-t\Delta_f})=\int_{\C^n}\tr_{\C^n}p(z,z,t;u)\emph{dvol}_h(z).\footnote{We emphasize that $p(z,z,t)$ is a matrix in the internal indices, $\tr_{\C^n}$ denote the trace over these indices.}$$
\epr

There is another way to consider the trace of $e^{-t\Delta_{f_0}}$ and $e^{-t\Delta_f}$. Recall that $\Delta_{f_0}$ and $\Delta_f$ both have purely discrete spectrum acting on the $L^2$-integrable form space, so do $e^{-t\Delta_{f_0}}$ and $e^{-t\Delta_f}$.
Let $\{\lambda_i\}$ represent the spectra of $\Delta_{f_0}$ or $\Delta_{f}$,  Fan-Fang proved that $e^{-t\Delta_{f_0}}$ and $e^{-t\Delta_f}$ are trace-class operators via $\sum_{i=1}^{\infty}e^{-t\lambda_i}<\infty$, which is done mainly by some discussions about compact perturbation of operators, and comparing the spectra of $\Delta_{f_0}^0$ with those of $\Delta_n^0$, where $\Delta_n^0:=\Delta_{\frac{1}{2}(z_1^2+\cdots+z_n^2)}^0$ is associated to $n$-harmonic oscillator.

\vskip 0.1cm
\noindent\textbf{Remark:} (1) It is worth to mention that Fan-Fang's proof starts from the spectra without the condition $q_M-q_m<\frac{1}{3}$.
\vskip 0.1cm
(2) In \cite{Fan2016Torsion}, Fan-Fang also proved the heat trace result in the case $X=(\C^*)^n$, $f_0$ is a non-degenerate and convenient Laurent polynomial, which is very important in the LG B-model as well.

\subsubsection{Zeta function and vanishing theorem}
\bd Under the weight condition $(\star)$ for $f_0$ and $f$, there exists a constant $C_f>0$ such that for $\re(s)>C_f$, the $i$-th zeta function associated to $f$ is defined to be:

$$\zeta_{f}^{i}(s)=\frac{1}{2\Gamma(s)}\int_{0}^{\infty}t^{s-1}\sum_{k=0}^{2n}(-1)^kk^i\Tr(e^{-t\Delta_{f}^k}-\Pi_k)dt,$$
where $\Pi_k$ is the harmonic projection operator $\Pi_{k}: \mathcal{A}_{M}^k(\C^n)\rightarrow \H^k$.% $\Pi_{0k}:
\ed

Immediately, we know that $\zeta_f^0(s)=0$, since $\sum_{k=0}^{2n}\Tr(-1)^k(e^{-t\Delta_f^k}-\Pi_k)=0$ by the Hodge decomposition. $\sum_{k=0}^{2n}\Tr(-1)^ke^{-t\Delta_f^k}$ is also well known as the Witten index in the physical literature. Furthermore, we consider the case $i=1$. In \cite{Fan2016Torsion}, Fan-Fang proved the following vanishing result for $\zeta_{f_0}^1(s)$.
\bl [Theorem 6.9 in \cite{Fan2016Torsion}] \label{vanish0}
Let $f_0$ be a non-degenerate and quasi-homogeneous polynomial on $\C^n$. Then
$$\zeta_{f_0}^1(s)=0.$$
\el
\bp For simplicity, define $Q_{f_0}=\sum_{k=0}^{2n}\Tr(-1)^kk(e^{-t\Delta_{f_0}^k}-\Pi_{0k})$, where $\Pi_{0k}: L^2\mathcal{A}^k\rightarrow \H_0^k$ is the harmonic projection.

On the one hand, the hodge star operator $*$ provides the relation
$$*\Delta_{f_0}=\Delta_{-f_0}*.$$
More precisely, $*\Delta_{f_0}^k=\Delta_{-f_0}^{2n-k}*$, for $k=0,1,\ldots,2n$. Then we have $$\Tr(e^{-t\Delta_{f_0}^k})=\Tr(e^{-t\Delta_{-f_0}^{2n-k}}).$$
Let $\Pi_{0k}'$ be the projection to the subspace $\ker(\Delta_{-f_0}\big|_{L^2\mathcal{A}^k})$. By Theorem \ref{dimH}, we have $$\Tr(-1)^kk^i\Pi_{0k}=(-1)^kk^i\mu\delta_{nk},\quad \Tr(-1)^kk^i\Pi_{0k}'=(-1)^kk^i\mu\delta_{nk}.$$
Therefore, we can compute $Q_{f_0}$ as
\begin{equation}\label{Q0}
\begin{aligned}
Q_{f_0}=~&\sum_{k=0}^{2n}\Tr(-1)^kk(e^{-t\Delta_{f_0}^k}-\Pi_{0k})\\
       =~&\sum_{k=0}^{2n}\Tr(-1)^kke^{-t\Delta_{-f_0}^{2n-k}}-\sum_{k=0}^{2n}\Tr(-1)^kk\Pi_{0k}\\
       =~&\sum_{k=0}^{2n}\Tr(-1)^{2n-k}(2n-k)e^{-t\Delta_{-f_0}^k}-(-1)^nn\mu\\
       =~&2n\Tr(-1)^ke^{-t\Delta_{-f_0}^k}-\sum_{k=0}^{2n}\Tr(-1)^{k}k(e^{-t\Delta_{-f_0}^k}-\Pi_{0k}')-2(-1)^nn\mu\\
       =~&2n\left[\sum_{k=0}^{2n}\Tr(-1)^ke^{-t\Delta_{-f_0}^k}-(-1)^n\mu\right]-Q_{-f_0}=-Q_{-f_0}.
\end{aligned}
\end{equation}

On the other hand, since $f_0$ is quasi-homogeneous, there exist $d,k_1,\ldots,k_n\in\mathbb{N}$ such that for any $\lambda\in\C^*$,
$$\lambda^d f_0(z_1,\ldots,z_n)=f_0(\lambda^{k_1}z_1,\ldots,\lambda^{k_n}z_n).$$
Hence, we can choose $\xi$ such that $\xi^d=-1$. Consider the scaling $I_{\xi}(z_i)=\xi^{k_i}z_i$, for $i=1,\ldots,n$. Then we have $f_0(I_{\xi}z)=-f_0(z)$ and the pull back $I_{\xi}^*: \mathcal{A}^*(\C^n)\rightarrow \mathcal{A}^*(\C^n)$. It is direct to check that
$$(I_{\xi}^*)^{-1}\circ\Delta_{-f_0}\circ I_{\xi}^*=\Delta_{f_0}.$$
More precisely, $(I_{\xi}^*)^{-1}\circ\Delta_{-f_0}^k\circ I_{\xi}^*=\Delta_{f_0}^k.$
Similarly, one can prove that $Q_{f_0}=Q_{-f_0}$. Combining it with \eqref{Q0}, we obtain $Q_{f_0}=0$. Therefore,
we get $\zeta_{f_0}^1(s)=0$.
\ep

\noindent\textbf{Remark:} (1) In \cite{Fan2016Torsion}, the authors prove that $e^{-t\Delta_{f_0}}$ is trace-class via studying their spectra when $f_0$ is a non-degenerate and quasi-homogeneous polynomial. So $\zeta_{f_0}^1(s)$ holds without the weight condition $q_M-q_m<\frac{1}{3}$.

(2) The proof here doesn't depend on the constant positive-definite metric on $\C^n$, that is, if we choose another constant positive-definite metric $h_1$ on $\C^n$, we have $\zeta_{f_0}(s;h_1)=0$. In fact, let $h_0$ denote the standard metric on $\C^n$. Set $h_c=(1-c)h_0+ch_1$, $0\leq h\leq1$, and $\ast$ is short for the corresponding Hodge star operator $\ast_c$. Since $\ast^2=(-1)^k$, we have $\dot{\ast}\ast^{-1}=-\ast^{-1}\dot{\ast}=-\a$.
Then
$$\frac{\p}{\p c}\Delta_{f_0}=\bar{\p}_{f_0}\a\bar{\p}_{f_0}^{\dag}-\bar{\p}_{f_0}\bar{\p}_{f_0}^{\dag}\a+\a\bar{\p}_{f_0}^{\dag}\bar{\p}_{f_0}-\bar{\p}_{f_0}^{\dag}\a\bar{\p}_{f_0}$$
By Duhamel formual and Equations \eqref{N0}, we have
\begin{flalign*}
&\frac{\p}{\p c}\Tr(-1)^NNe^{-t\Delta_{f_0}}\\
=&-t\Tr(-1)^NN\left(\bar{\p}_{f_0}\a\bar{\p}_{f_0}^{\dag}-\bar{\p}_{f_0}\bar{\p}_{f_0}^{\dag}\a+\a\bar{\p}_{f_0}^{\dag}\bar{\p}_{f_0}-\bar{\p}_{f_0}^{\dag}\a\bar{\p}_{f_0}\right)e^{-t\Delta_{f_0}}\\
=&-t\Tr(-1)^N\left(\bar{\p}_{f_0}\a\bar{\p}_{f_0}^{\dag}+\bar{\p}_fN\a\bar{\p}_{f_0}^{\dag}-N\bar{\p}_{f_0}\bar{\p}_{f_0}^{\dag}\a+N\a\bar{\p}_{f_0}^{\dag}\bar{\p}_{f_0}+\bar{\p}_{f_0}^{\dag}\a\bar{\p}_{f_0}-\bar{\p}_{f_0}^{\dag}N\a\bar{\p}_{f_0}\right)e^{-t\Delta_{f_0}}\\
=~&t\Tr(-1)^N\a\Delta_fe^{-t\Delta_{f_0}}=-t\frac{d}{dt}\Tr(-1)^N\a e^{-t\Delta_{f_0}}.
\end{flalign*}
Using the same discussion in the proof above, we have
$$\zeta_{f_0}^1(s;h_c)=-\zeta_{f_0}^1(s;h_c),\quad \zeta_{f_0}^1(s;h_c)=\zeta_{f_0}^1(s;h_c).$$

\vskip 0.1cm
Now we prove that the vanishing theorem also holds for the deformed case under the weight condition $(\star)$.

\bt \label{vanishf} Under the weight condition $(\star)$ for $f_0$ and $f$,

$$\zeta_{f}^1(s)=0.$$
\et
Before proving the theorem, let us rewrite the $i$-th zeta function in a more convenient way via the number operator $N$;
\begin{equation}\label{zeta}
\zeta_{f}^i(s)=\frac{1}{2\Gamma(s)}\int_{0}^{\infty}t^{s-1}\Tr(-1)^NN^i(e^{-t\Delta_{f}}-\Pi)dt,
\end{equation}
where $\Pi: L^2\mathcal{A}_{M}^*(\C^n)\rightarrow\H^n$ is the harmonic projection as before.

The following two sets of relations will be used repeatedly in the sequel.

\begin{equation}\label{N}
\begin{aligned}
&\{\bar{\p}_f,\p_f\}=\{\bar{\p}_f,\p_f^{\dag}\}=\{\bar{\p}_f^{\dag},\p_f\}=\{\bar{\p}_f^{\dag},\p_f^{\dag}\}=0,\\
&[N,\bar{\p}_f]=\bar{\p}_f, [N,\bar{\p}_f^{\dag}]=-\bar{\p}_f^{\dag},\\
&[N,\p_f]=\p_f, [N,\p_f^{\dag}]=-\p_f^{\dag},
\end{aligned}
\end{equation}
and
\begin{equation}\label{ij}
\begin{aligned}
&[\p_i, \bar{\p}_f]=[\p_f, \phi_i], [\p_i, \p_f^{\dag}]=-[\bar{\p}_f^{\dag},\phi_i],\\
&[\bap_{\bar{j}},\p_f]=[\bar{\p}_f,\bar{\phi}_{\bj}], [\bap_{\bar{j}}, \bar{\p}_f^{\dag}]=-[\p_f^{\dag},\bar{\phi}_{\bj}],\\
&[\bar{\p}_f, \phi_i]=0, [\p_f^{\dag},\phi_i]=0,\\
&[\p_f,\bar{\phi}_{\bj}]=0, [\bar{\p}_f^{\dag},\bar{\phi}_{\bj}]=0.
\end{aligned}
\end{equation}

\bl[Duhamel Formula]\label{Duhamelformula} For the heat operator $e^{-t\Delta_f}$, we have
$$\p_ie^{-t\Delta_f}=\int_0^te^{-t'\Delta_f}(-\p_i\Delta_f)e^{-(t-t')\Delta_f}dt',\quad \bap_{\bar{j}}e^{-t\Delta_f}=\int_0^te^{-t'\Delta_f}(-\bap_{\bar{j}}\Delta_f)e^{-(t-t')\Delta_f}dt'.$$
\el

\bl \label{AB}
For the Hilbert-Schmidt operators $A,B,O$ and the number operator $N$, assume that they satisfy one of the following relations
\begin{enumerate}
  \item $\{(-1)^{N},A\}=0, [A,O]=0;$
  \item $\{(-1)^{N},B\}=0, [B,O]=0.$
\end{enumerate}
Then
$$\Tr(-1)^{N}\{A,B\}O=0.$$
\el
\bp
Suppose $\{(-1)^{N},A\}=0, [A,O]=0$, then
\begin{flalign*}
\Tr(-1)^{N}ABO~=~&\Tr BO(-1)^{N}A~=~\Tr BO(-A)(-1)^{N}\\
               =~&-\Tr BAO(-1)^{N}=-\Tr(-1)^{N}BAO.
\end{flalign*}
The second case is similar.
\ep
\noindent\textbf{Remark}: This lemma is called the $AB$ argument in physics, and frequently comes into the discussion of the variation of some new supersymmetry index; see, for example, in \cite{Cecotti1992A}.\\

Set $I=(I_1,\ldots,I_n), J=(J_1,\ldots,J_n)\in\mathbb{N}^n$, $K=(K_1,\ldots,K_s), L=(L_1,\ldots,L_s)\in\mathbb{N}^s$, and furthermore, $$\p_z^I\bap_{\bar{z}}^J=\p_{z_1}^{I_1}\cdots\p_{z_n}^{I_n}\bap_{\bar{z}_1}^{J_1}\cdots\bap_{\bar{z}_n}^{J_n},\quad \phi^K\bar{\phi}^L=\phi_1^{K_1}\cdots\phi_s^{K_s}\bar{\phi}_{\bar{1}}^{L_1}\cdots\bar{\phi}_{\bar{s}}^{L_s}.$$
Then we have

\bl\label{HS} Under the weight condition $(\star)$ for $f_0$ and $f$, the operators of the form $\p_{z}^I\bap_{\bar{z}}^J\phi^K\bar{\phi}^Le^{-t\Delta_f}$ are Hilbert-Schmidt.
\el
\bp Let $p(z,w,t;u)$ be the kernel function of $e^{-t\Delta_f}$, then the kernel functions for the operators $\p_{z_{\nu}}e^{-t\Delta_f}$ and $\phi_ie^{-t\Delta_f}$ are given by $\p_{z_{\nu}}p(z,w,t;u)$ and $\phi_i(z)p(z,w,t;u)$ respectively. It is an easy observation from the construction of $p(z,w,t;u)$ that these kernel functions are $L^2$-integrable, therefore, the corresponding operators are Hilbert-Schmidt. So do the operators of the form $\p_{z}^I\bap_{\bar{z}}^J\phi^K\bar{\phi}^Le^{-t\Delta_f}$.
\ep

\bp[Proof of Theorem \ref{vanishf}] Note that $e^{-t\Delta_f}$ depends smoothly on the parameters $u^i, \bar{u}^{\bar{j}}$. And when $u\in M$, we have $\Tr(-1)^NN\Pi=(-1)^nn\mu$. Therefore, $\Tr(-1)^NN(e^{-t\Delta_f}-\Pi)$ is differentiable with respect to $u^i$, $\bar{u}^{\bar{j}}$.
Then to prove Theorem \ref{vanishf}, it suffices to prove that $\zeta_f^1(s)$ does not depend on the deformation parameter $u^i$, $\bar{u}^{\bar{j}}$. That is, to show
$$\p_i\zeta_f^1(s)=0,\quad \bar{\p}_{\bar{j}}\zeta_f^1(s)=0.$$

Let us compute the derivative with respect to $u^i$, $\bar{u}^{\bar{j}}$ on the super-trace first. By equations \eqref{N}, \eqref{ij} and Lemmas \ref{Duhamelformula} \ref{AB}, \ref{HS}, we have
\begin{equation*}
\begin{aligned}
\p_i\Tr(-1)^NN(e^{-t\Delta_f}-\Pi)~=~&-\Tr(-1)^NN\int_{0}^te^{-t'\Delta_f}(\p_i\Delta_f)e^{-(t-t')\Delta_f}dt'\\
                                   \overset{(a)}=~&-t\Tr(-1)^NN(\p_i\Delta_f)e^{-t\Delta_f}\\
                                   \overset{(b)}=~&-t\Tr(-1)^NN\p_i\{\bar{\p}_f^{\dag}, \bar{\p}_f\}e^{-t\Delta_f}\\
                                   =~&-t\Tr(-1)^NN\left(\bar{\p}_f^{\dag}[\p_i,\bar{\p}_f]+[\p_i,\bar{\p}_f]\bar{\p}_f^{\dag}\right)e^{-t\Delta_f}\\
                                   =~&-t\Tr(-1)^NN\left(\bar{\p}_f^{\dag}[\p_f,\phi_i]+[\p_f,\phi_i]\bar{\p}_f^{\dag}\right)e^{-t\Delta_f}\\
                                   =~&-t\Tr(-1)^N\left([N,\bar{\p}_f^{\dag}][\p_f,\phi_i]+\bar{\p}_f^{\dag}N[\p_f,\phi_i]+N[\p_f,\phi_i]\bar{\p}_f^{\dag}\right)e^{-t\Delta_f}\\
                                   \overset{(c)}=~&-t\Tr(-1)^N\left(-\bar{\p}_f^{\dag}[\p_f,\phi_i]\right)e^{-t\Delta_f}\\
                                   =~&-t\Tr(-1)^N\left(-\bar{\p}_f^{\dag}\p_f\phi_i+\bar{\p}_f^{\dag}\phi_i\p_f\right)e^{-t\Delta_f}\\
                                   \overset{(d)}=~&-t\Tr(-1)^N\left(-\bar{\p}_f^{\dag}\p_f\phi_i-\p_f\bar{\p}_f^{\dag}\phi_i\right)e^{-t\Delta_f}=0.
\end{aligned}
\end{equation*}
The equality $(a)$: It holds by the acyclic property of trace operator.
\vskip 0.1cm
\noindent The equality $(b)$: The two operators $e^{-t'\Delta_f}$ and $(\p_i\Delta_f)e^{-(t-t')\Delta_f}$ are both Hilbert-Schmidt operators, $N$ is a bounded operator, and $[N, e^{-t'\Delta_f}]=0$, then it follows from the acyclic property of trace.
\vskip 0.1cm
\noindent The equality $(c)$: Naively, it follows from the AB argument. But $\bar{\p}_f^{\dag}$ is an unbounded operator. This equality holds by
\begin{flalign*}
 &\Tr(-1)^N\left(\bar{\p}_f^{\dag}N[\p_f,\phi_i]+N[\p_f,\phi_i]\bar{\p}_f^{\dag}\right)e^{-t\Delta_f}\\
=~&\Tr(-1)^N\left(\bar{\p}_f^{\dag}N[\p_f,\phi_i]+N[\p_f,\phi_i]\bar{\p}_f^{\dag}\right)e^{-\frac{t}{2}\Delta_f}e^{-\frac{t}{2}\Delta_f}\\
=~&\Tr(-1)^N\bar{\p}_f^{\dag}e^{-\frac{t}{2}\Delta_f}N[\p_f,\phi_i]e^{-\frac{t}{2}\Delta_f}+\Tr(-1)^NN[\p_f,\phi_i]\bar{\p}_f^{\dag}e^{-t\Delta_f}\\
=~&-\Tr(-1)^NN[\p_f,\phi_i]e^{-\frac{t}{2}\Delta_f}\bar{\p}_f^{\dag}e^{-\frac{t}{2}\Delta_f}+\Tr(-1)^NN[\p_f,\phi_i]\bar{\p}_f^{\dag}e^{-t\Delta_f}\\
=~&-\Tr(-1)^NN[\p_f,\phi_i]\bar{\p}_f^{\dag}e^{-t\Delta_f}+\Tr(-1)^NN[\p_f,\phi_i]\bar{\p}_f^{\dag}e^{-t\Delta_f}=0.
\end{flalign*}
The equality $(d)$: We want to use the acyclic property of trace to move the operator $\p_f$ to the front. The process is similar to the equality $(b)$ that we put the operator $e^{-\frac{t}{2}\Delta_f}$ and the operator $\p_f$ together to move around the trace. In the next section, we will use this method repeatedly.

Therefore, we have $\p_i\zeta_f^1(s)=0$. Analogously, $\bap_{\bar{j}}\Tr(-1)^NN(e^{-t\Delta_f}-\Pi)=0$, and therefore $\bap_{\bar{j}}\zeta_f^1(s)=0$.
\ep

\noindent\textbf{Remarks:}
(1) Note that if we scale $f$ to $\b f$, $\b\in\C^*$, the heat kernel of $e^{-t\Delta_{\b f}}$ still existence under the weight condition $(\star)$, and smoothly depends on $\b\in\C^*$. It is believed that $\zeta_{\b f}^1(s)=0$. Indeed,
$$\frac{\p}{\p \b}\Delta_{\b f}=\{[\p_{\b},\bar{\p}_{\b f}],\bar{\p}_{\b f}^{\dag}\}=\bar{\p}_{\b f}^{\dag}[\p_{\b f},f]+[\p_{\b f}, f]\bar{\p}_{\b f}^{\dag},$$
the rest proceed as the proof of Theorem \ref{vanishf}.

(2) This theorem could be viewed as a counterpart of the vanishing theorem \cite{Ray1973Analytic} of the Ray-Singer analytic torsion, which also vanishes in the supertrace level
$$\sum_{p,q}\Tr(-1)^{p+q}p(e^{-t\Delta^{p,q}}-\Pi_{p,q})=0.$$
Furthermore, the symmetry of $p,q$ allows us to obtain
$$\sum_{p,q}\Tr(-1)^{p+q}(p+q)(e^{-t\Delta^{p,q}}-\Pi_{p,q})=0.$$

(3) In the 2d LG B-model, Cecotti, Fendley, Intriligator and Vafa \cite{Cecotti1992A} introduced some quantities which are of the form
$$I_l(\beta)=\Tr(-1)^FF^le^{-\beta H}, \text{ for } l\geq 0.$$
When $l=0$, it is the famous Witten index. For $l=1$, they proved that $I_1$ depends only on the F-term in the action functional, or in other words, it depends only on the superpotential $f$. It is for this reason that they regarded it as the supersymmetric index. Actually, it depends on the boundary condition at spatial infinity and they showed that it is an anti-symmetric matrix. This index is very important since it works as a bridge connecting the differential equation ($tt^*$ equation) and the integral equation (TBA equation). The surprising equivalence between the coupled integral equation and certain differential equations has not been proven mathematically. For $l\geq 2$, $I_l$ does depend on the D-term.\\

Because of the vanishing theorem, the next zeta function which interests us is
$$\zeta_f^2(s)=\frac{1}{2\Gamma(s)}\int_{0}^{\infty}t^{s-1}\Tr(-1)^NN^2(e^{-t\Delta_f}-\Pi)dt.$$
As we mentioned in the Introduction, on the CY B-side, the BCOV torsion could be interpreted as the genus 1 partition function of Calabi-Yau 3-fold target theory. The corresponding zeta function, denoted by $\zeta_{BCOV}(s)$, is
$$\zeta_{BCOV}(s)=\frac{1}{\Gamma(s)}\int_{0}^{\infty}t^{s-1}\Tr(-1)^{p+q}pq(e^{-t\Delta^{p,q}}-\Pi_{p,q})dt.$$
Because of the vanishing theorem in the complex geometry, it can be written as
$$\zeta_{BCOV}(s)=\frac{1}{2\Gamma(s)}\int_{0}^{\infty}t^{s-1}\Tr(-1)^{p+q}(p+q)^2(e^{-t\Delta^{p,q}}-\Pi_{p,q})dt.$$
It looks similar to our 2nd zeta function. So it is natural to hope that the 2nd torsion invariant could be related to the genus 1 partition function on the LG B-side in the spirit of the LG/CY correspondence. \\

\section{Computation of $\bap_{\bj}\p_i\Tr(-1)^NN^2(e^{-t\Delta_f}-\Pi)$}

The goal of this section is to obtain the transgression formula for $$\bar{\p}_{\bar{j}}\p_i\Tr(-1)^{N}N^2(e^{-t\Delta_f}-\Pi).$$
In order to compute it, we will again make full use of equations \eqref{N}, \eqref{ij} and Lemmas \ref{Duhamelformula}, \ref{AB}, \ref{HS}.
\vskip 0.1cm
Note that the following identity is given in the proof of Theorem \ref{vanishf}.
\bl\label{1}
$\Tr(-1)^N\bar{\p}_{f}^{\dag}[\p_f,\phi_i]e^{-t\Delta_f}=0$, $\Tr(-1)^N\p_{f}^{\dag}[\bar{\p}_f,\bar{\phi}_{\bar{j}}]e^{-t\Delta_f}=0$.
\el

Now let us first compute $\p_i\Tr(-1)^NN^2(e^{-t\Delta_f}-\Pi)$, the result is given by
\bl $\p_i\Tr(-1)^NN^2(e^{-t\Delta_f}-\Pi)=2t\Tr(-1)^NN\bar{\p}_f^{\dag}[\p_f,\phi_i]e^{-t\Delta_f}.$
\el
\bp By equations \eqref{N}, \eqref{ij} and Lemmas \ref{AB}, \ref{HS}, \ref{1}, we have
\begin{flalign*}
  &\p_{i}\Tr(-1)^{N}N^2(e^{-t\Delta_f}-\Pi)\\
\overset{(a)}=~&-t\Tr(-1)^{N}N^2\left(\bar{\p}_f^{\dag}[\p_i,\bar{\p}_f]+[\p_i,\bar{\p}_f]\bar{\p}_f^{\dag}\right)e^{-t\Delta_f}\\
=~&-t\Tr(-1)^{N}N^2\left(\bar{\p}_f^{\dag}[\p_f,\phi_i]+[\p_f,\phi_i]\bar{\p}_f^{\dag}\right)e^{-t\Delta_f}\\
=~&-t\Tr(-1)^NN\left([N,\bar{\p}_f^{\dag}][\p_f,\phi_i]+\bar{\p}_f^{\dag}N[\p_f,\phi_i]+N[\p_f,\phi_i]\bar{\p}_f^{\dag}\right)e^{-t\Delta_f}\\
=~&-t\left\{-\Tr(-1)^NN\bar{\p}_f^{\dag}[\p_f,\phi_i]e^{-t\Delta_f}+\Tr(-1)^N[N,\bar{\p}_f^{\dag}]N[\p_f,\phi_i]e^{-t\Delta_f}\right.\\
 ~&\left.+\Tr(-1)^N\bar{\p}_f^{\dag}N^2[\p_f,\phi_i]e^{-t\Delta_f}+\Tr(-1)^NN^2[\p_f,\phi_i]\bar{\p}_f^{\dag} e^{-t\Delta_f}\right\}\\
\overset{(b)}=~&t\Tr(-1)^NN\bar{\p}_f^{\dag}[\p_f,\phi_i]e^{-t\Delta_f}+t\Tr(-1)^N\bar{\p}_f^{\dag}N[\p_f,\phi_i]e^{-t\Delta_f}\\
=~&2t\Tr(-1)^NN\bar{\p}_f^{\dag}[\p_f,\phi_i]e^{-t\Delta_f}.
\end{flalign*}
The equalities $(a), (b)$ are similar to the computation of $\p_i\Tr(-1)^NN(e^{-t\Delta_f}-\Pi)$, in what follows, we will omit the discussions. The last equality is given by Lemma \ref{1}.
\ep

Next, let us compute $\bap_{\bj}\p_i\Tr(-1)^NN^2(e^{-t\Delta_f}-\Pi)$:
\begin{flalign*}
  &\bap_{\bj}\p_i\Tr(-1)^NN^2(e^{-t\Delta_f}-\Pi)\\
=~&2t\Tr(-1)^NN[\bap_{\bj},\bar{\p}_f^{\dag}][\p_f,\phi_i]e^{-t\Delta_f}+2t\Tr(-1)^NN\bar{\p}_f^{\dag}[\p_f,\phi_i][\bap_{\bj},e^{-t\Delta_f}].
\end{flalign*}
For convenience, denote
\begin{flalign*}
A_1&=\Tr(-1)^NN[\bap_{\bj},\bar{\p}_f^{\dag}][\p_f,\phi_i]e^{-t\Delta_f},\\
A_2&=\Tr(-1)^NN\bar{\p}_f^{\dag}[\p_f,\phi_i][\bap_{\bj},e^{-t\Delta_f}].
\end{flalign*}
Then we have
\bl\label{A1}
 $A_1=\Tr(-1)^N(\Delta_f\bar{\phi}_{\bar{j}}\phi_i+N\bar{\phi}_{\bar{j}}\Delta_f\phi_i-N\bar{\phi}_{\bar{j}}\phi_i\Delta_f)e^{-t\Delta_f}$.
\el
\bp By equations \eqref{N}, \eqref{ij}, and Lemmas \ref{AB}, \ref{HS}, we have
\begin{flalign*}
  &\Tr(-1)^NN[\bap_{\bj},\bar{\p}_f^{\dag}][\p_f,\phi_i]e^{-t\Delta_f}\\
=~&-\Tr(-1)^NN[\p_f^{\dag},\bar{\phi}_{\bar{j}}][\p_f,\phi_i]e^{-t\Delta_f}\\
=~&-\Tr(-1)^NN\left(\p_f^{\dag}\bar{\phi}_{\bar{j}}\p_f\phi_i-\p_f^{\dag}\bar{\phi}_{\bar{j}}\phi_i\p_f-\bar{\phi}_{\bar{j}}\p_f^{\dag}\p_f\phi_i+\bar{\phi}_{\bar{j}}\p_f^{\dag}\phi_i\p_f\right)e^{-t\Delta_f}\\
=~&-\Tr(-1)^N\left(-\p_f^{\dag}\bar{\phi}_{\bar{j}}\p_f\phi_i+\p_f^{\dag}N\bar{\phi}_{\bar{j}}\p_f\phi_i+\p_f^{\dag}\bar{\phi}_{\bar{j}}\phi_i\p_f-\p_f^{\dag}N\bar{\phi}_{\bar{j}}\phi_i\p_f\right.\\
 ~&\left.\quad\quad\quad\quad\quad-N\bar{\phi}_{\bar{j}}\p_f^{\dag}\p_f\phi_i+N\bar{\phi}_{\bar{j}}\phi_i\p_f^{\dag}\p_f \right)e^{-t\Delta_f}\\
=~&-\Tr(-1)^N\left(-\p_f^{\dag}\p_f\bar{\phi}_{\bar{j}}\phi_i-\p_f\p_f^{\dag}\bar{\phi}_{\bar{j}}\phi_i-N\bar{\phi}_{\bar{j}}\p_f\p_f^{\dag}\phi_i+N\bar{\phi}_{\bar{j}}\phi_i\p_f\p_f^{\dag}\right.\\
 ~&\left.\quad\quad\quad\quad\quad-N\bar{\phi}_{\bar{j}}\p_f^{\dag}\p_f\phi_i+N\bar{\phi}_{\bar{j}}\phi_i\p_f^{\dag}\p_f \right)e^{-t\Delta_f}\\
=~&\Tr(-1)^N\left(\Delta_f\bar{\phi}_{\bar{j}}\phi_i+N\bar{\phi}_{\bar{j}}\Delta_f\phi_i-N\bar{\phi}_{\bar{j}}\phi_i\Delta_f\right)e^{-t\Delta_f}.\qedhere
\end{flalign*}
\ep

The computation of $A_2$ is a little more complicated, since it involves more noncommutative relations.
\begin{flalign*}
A_2=~&-\Tr(-1)^NN\bar{\p}_f^{\dag}[\p_f,\phi_i]\int_0^te^{-t'\Delta_f}(\p_f^{\dag}[\bap_{\bar{j}},\p_f]+[\bap_{\bar{j}},\p_f]\p_f^{\dag})e^{-(t-t')\Delta_f}dt'\\
   =~&-\Tr(-1)^N\bar{\p}_f^{\dag}[\p_f,\phi_i]\int_0^te^{-t'\Delta_f}(-\p_f^{\dag}[\bap_{\bar{j}},\p_f]+\p_f^{\dag}N[\bap_{\bar{j}},\p_f]+N[\bap_{\bar{j}},\p_f]\p_f^{\dag})e^{-(t-t')\Delta_f}dt'\\
   =~&\Tr(-1)^N(\bar{\p}_f^{\dag}\p_f\phi_i-\bar{\p}_f^{\dag}\phi_i\p_f)\p_f^{\dag}\int_0^te^{-t'\Delta_f}[\bar{\p}_f,\bar{\phi}_{\bar{j}}]e^{-(t-t')\Delta_f}dt'\\
    ~&-\Tr(-1)^N(\bar{\p}_f^{\dag}\p_f\phi_i-\bar{\p}_f^{\dag}\phi_i\p_f)\p_f^{\dag}\int_0^te^{-t'\Delta_f}N[\bar{\p}_f,\bar{\phi}_{\bar{j}}]e^{-(t-t')\Delta_f}dt'\\
    ~&+\Tr(-1)^N\p_f^{\dag}(\bar{\p}_f^{\dag}\p_f\phi_i-\bar{\p}_f^{\dag}\phi_i\p_f)\int_0^te^{-t'\Delta_f}N[\bar{\p}_f,\bar{\phi}_{\bar{j}}]e^{-(t-t')\Delta_f}dt'\\
   =~&-\Tr(-1)^N\bar{\p}_f^{\dag}\p_f^{\dag}[\p_f, \phi_i]\int_0^te^{-t'\Delta_f}[\bar{\p}_f,\bar{\phi}_{\bar{j}}]e^{-(t-t')\Delta_f}dt'\\
    ~&+\Tr(-1)^N\bar{\p}_f^{\dag}[\Delta_f, \phi_i]\int_0^te^{-t'\Delta_f}[\bar{\p}_f,\bar{\phi}_{\bar{j}}]e^{-(t-t')\Delta_f}dt'\\
    ~&-\Tr(-1)^N\bar{\p}_f^{\dag}[\Delta_f, \phi_i]\int_0^te^{-t'\Delta_f}N[\bar{\p}_f,\bar{\phi}_{\bar{j}}]e^{-(t-t')\Delta_f}dt'.
\end{flalign*}

Again, for convenience, denote
\begin{flalign*}
&B_1=-\Tr(-1)^N\bar{\p}_f^{\dag}\p_f^{\dag}[\p_f, \phi_i]\int_0^te^{-t'\Delta_f}[\bar{\p}_f,\bar{\phi}_{\bar{j}}]e^{-(t-t')\Delta_f}dt',\\
&B_2=\Tr(-1)^N\bar{\p}_f^{\dag}[\Delta_f, \phi_i]\int_0^te^{-t'\Delta_f}[\bar{\p}_f,\bar{\phi}_{\bar{j}}]e^{-(t-t')\Delta_f}dt',\\
&B_3=-\Tr(-1)^N\bar{\p}_f^{\dag}[\Delta_f, \phi_i]\int_0^te^{-t'\Delta_f}N[\bar{\p}_f,\bar{\phi}_{\bar{j}}]e^{-(t-t')\Delta_f}dt'.
\end{flalign*}

\noindent Then we have
\bl \label{B1}
 $B_1=-\Tr(-1)^N\int_0^te^{-t'\Delta_f}\bar{\phi}_{\bar{j}}e^{-(t-t')\Delta_f}(\Delta_f)^2\phi_idt'$.
\el
\bp By equations \eqref{N}, \eqref{ij} and Lemma \ref{HS}, we have
\begin{flalign*}
B_1=~&-\Tr(-1)^N\bar{\p}_f^{\dag}\p_f^{\dag}(\p_f\phi_i-\phi_i\p_f)\int_0^te^{-t'\Delta_f}(\bar{\p}_f\bar{\phi}_{\bar{j}}-\bar{\phi}_{\bar{j}}\bar{\p}_f)e^{-(t-t')\Delta_f}dt'\\
   =~&-\Tr(-1)^N\bar{\p}_f^{\dag}\p_f^{\dag}(\p_f\phi_i-\phi_i\p_f)\bar{\p}_f\int_0^te^{-t'\Delta_f}\bar{\phi}_{\bar{j}}e^{-(t-t')\Delta_f}dt'\\
    ~&-\Tr(-1)^N\bar{\p}_f\bar{\p}_f^{\dag}\p_f^{\dag}(\p_f\phi_i-\phi_i\p_f)\int_0^te^{-t'\Delta_f}\bar{\phi}_{\bar{j}}e^{-(t-t')\Delta_f}dt'\\
   =~&-\Tr(-1)^N(\Delta_f\p_f^{\dag}\p_f\phi_i-\Delta_f\p_f^{\dag}\phi_i\p_f)\int_0^te^{-t'\Delta_f}\bar{\phi}_{\bar{j}}e^{-(t-t')\Delta_f}dt'\\
   \overset{(a)}=~&-\Tr(-1)^N\Delta_f(\p_f^{\dag}\p_f\phi_i+\p_f\p_f^{\dag}\phi_i)\int_0^te^{-t'\Delta_f}\bar{\phi}_{\bar{j}}e^{-(t-t')\Delta_f}dt'\\
   =~&-\Tr(-1)^N(\Delta_f)^2\phi_i\int_0^te^{-t'\Delta_f}\bar{\phi}_{\bar{j}}e^{-(t-t')\Delta_f}dt'\\
   =~&-\Tr(-1)^N\int_0^te^{-t'\Delta_f}\bar{\phi}_{\bar{j}}e^{-(t-t')\Delta_f}(\Delta_f)^2\phi_idt'.
\end{flalign*}
In the equality $(a)$, we use the relation $[\p_f,\bar{\phi}_{\bar{j}}]=0$, and
\begin{equation*}
\int_0^te^{-t'\Delta_f}\bar{\phi}_{\bar{j}}e^{-(t-t')\Delta_f}dt'=e^{-\frac{t}{2}\Delta_f}\int_0^{\frac{t}{2}}e^{-t'\Delta_f}\bar{\phi}_{\bar{j}}e^{-(\frac{t}{2}-t')\Delta_f}dt'+\int_0^{\frac{t}{2}}e^{-t'\Delta_f}\bar{\phi}_{\bar{j}}e^{-(\frac{t}{2}-t')\Delta_f}dt'e^{-\frac{t}{2}\Delta_f}
\end{equation*}
to move the operator $\p_f$ around the trace.
\ep

\noindent The methods to compute $B_2$ and $B_3$ are similar.
\bl\label{B23} The terms $B_2$ and $B_3$ are given by
\begin{flalign*}
B_2&=\Tr(-1)^N(\bar{\phi}_{\bar{j}}\Delta_f\phi_i-\Delta_f\phi_i\bar{\phi}_{\bar{j}})e^{-t\Delta_f},\\
B_3&=\Tr(-1)^N(\Delta_f\phi_i\bar{\phi}_{\bar{j}}-N\bar{\phi}_{\bar{j}}\Delta_f\phi_i+N\bar{\phi}_{\bar{j}}\phi_i\Delta_f)e^{-t\Delta_f}.
\end{flalign*}
\el
\bp By equations \eqref{N}, \eqref{ij}, and Lemma \ref{HS}, we see that
\begin{flalign*}
B_2=~&\Tr(-1)^N\bar{\p}_f^{\dag}\phi_i\int_0^te^{-t'\Delta_f}[\bar{\p}_f,\bar{\phi}_{\bar{j}}]e^{-(t-t')\Delta_f}\Delta_fdt'\\
    ~&-\Tr(-1)^N\bar{\p}_f^{\dag} \phi_i\int_0^t\Delta_fe^{-t'\Delta_f}[\bar{\p}_f,\bar{\phi}_{\bar{j}}]e^{-(t-t')\Delta_f}dt',\\
   =~&\Tr(-1)^N\bar{\p}_f^{\dag}\phi_i\int_0^t\frac{d}{dt'}\left(e^{-t'\Delta_f}[\bar{\p}_f,\bar{\phi}_{\bar{j}}]e^{-(t-t')\Delta_f}\right)dt'\\
   =~&\Tr(-1)^N\bar{\p}_f^{\dag}\phi_i\left(e^{-t\Delta_f}[\bar{\p}_f,\bar{\phi}_{\bar{j}}]-[\bar{\p}_f,\bar{\phi}_{\bar{j}}]e^{-t\Delta_f}\right)\\
   =~&\Tr(-1)^N\bar{\phi}_{\bar{j}}\bar{\p}_f^{\dag}\phi_i\bar{\p}_fe^{-t\Delta_f}+\Tr(-1)^N\bar{\phi}_{\bar{j}}\bar{\p}_f\bar{\p}_f^{\dag}\phi_ie^{-t\Delta_f}\\
    ~&-\Tr(-1)^N\bar{\p}_f^{\dag}\phi_i\bar{\p}_f\bar{\phi}_{\bar{j}}e^{-t\Delta_f}-\Tr(-1)^N\bar{\p}_f\bar{\p}_f^{\dag}\phi_i\bar{\phi}_{\bar{j}}e^{-t\Delta_f}\\
   =~&\Tr(-1)^N(\bar{\phi}_{\bar{j}}\Delta_f\phi_i-\Delta_f\phi_i\bar{\phi}_{\bar{j}})e^{-t\Delta_f}.
\end{flalign*}
Similarly,
\begin{flalign*}
B_3=~&-\Tr(-1)^N\bar{\p}_f^{\dag}\phi_i\int_0^te^{-t'\Delta_f}N[\bar{\p}_f,\bar{\phi}_{\bar{j}}]e^{-(t-t')}\Delta_fdt'\\
    ~&+\Tr(-1)^N\bar{\p}_f^{\dag} \phi_i\int_0^t\Delta_fe^{-t'\Delta_f}N[\bar{\p}_f,\bar{\phi}_{\bar{j}}]e^{-(t-t')}dt',\\
    =~&-\Tr(-1)^N\bar{\p}_f^{\dag}\phi_i\left(e^{-t\Delta_f}N[\bar{\p}_f,\bar{\phi}_{\bar{j}}]-N[\bar{\p}_f,\bar{\phi}_{\bar{j}}]e^{-t\Delta_f}\right)\\
    =~&\Tr(-1)^N(\Delta_f\phi_i\bar{\phi}_{\bar{j}}-N\bar{\phi}_{\bar{j}}\Delta_f\phi_i+N\bar{\phi}_{\bar{j}}\phi_i\Delta_f)e^{-t\Delta_f}.\qedhere
\end{flalign*}
\ep

By Lemmas \ref{A1}, \ref{B1}, \ref{B23}, we get
\begin{equation}\label{transgress1}
\begin{aligned}
\bap_{\bar{j}}\p_i\Tr(-1)^NN^2(e^{-t\Delta_f}-\Pi)=~&2t\Tr(-1)^N\Delta_f\phi_i\bar{\phi}_{\bar{j}}e^{-t\Delta_f}+2t\Tr(-1)^N\bar{\phi}_{\bar{j}}\Delta_f\phi_ie^{-t\Delta_f}\\
                                                   &-2t\Tr(-1)^N\int_0^te^{-t'\Delta_f}\bar{\phi}_{\bar{j}}.
\end{aligned}
\end{equation}

Furthermore, we see that the right hand side of the equation \eqref{transgress1} above admits a differential of some function with respect to $t$.

\bt[Transgression Formula]\label{Transgress} Under the weight condition $(\star)$ for $f_0$ and $f$,
\begin{equation}\label{transgress2}
\begin{aligned}
\bar{\p}_{\bar{j}}\p_i\Tr(-1)^NN^2(e^{-t\Delta_f}-\Pi)~=~&-2t\frac{d}{dt}\Tr(-1)^N\phi_i\bar{\phi}_{\bar{j}}e^{-t\Delta_f}\\
                                                        ~&+2t\frac{d}{dt}\Tr(-1)^N\int_0^te^{-t'\Delta_f}\bar{\phi}_{\bar{j}}e^{-(t-t')\Delta_f}\Delta_f\phi_idt'.
\end{aligned}
\end{equation}
\et

\noindent\textbf{Remark:} The equation \eqref{transgress2} looks like the transgression formula of the Chern character in \cite{SouleAnalytic}, so by analogy we call it the transgression formula associated to $\Delta_f$.\\

\section{Anomaly Formula}

\subsection{Meromorphic extension of the zeta function and singularity torsion}
Recall that our $i$-th zeta function associated to $\Delta_f$ is defined to be
$$\zeta_{f}^i(s)=\frac{1}{2\Gamma(s)}\int_{0}^{\infty}t^{s-1}\Tr(-1)^NN^i(e^{-t\Delta_{f}}-\Pi)dt\quad \text{for } \re(s)>C_f,$$
which is holomorphic with respect to $\re(s)>C_f$. In this subsection, we consider its meromorphic extension.

For simplicity, we drop off the subscript $f$, and denote $\zeta_{f}^i(s)$ by $\zeta^i(s)$. Let us split the integral over $t$ into $[0,1]$ and $[1,\infty)$, and write
\begin{flalign*}
&\zeta_1^i=\frac{1}{2\Gamma(s)}\int_{0}^{1}t^{s-1}\Tr(-1)^NN^i(e^{-t\Delta_{f}}-\Pi)dt,\\
&\zeta_2^i=\frac{1}{2\Gamma(s)}\int_{1}^{\infty}t^{s-1}\Tr(-1)^NN^i(e^{-t\Delta_{f}}-\Pi)dt.
\end{flalign*}

First, we have
\bl The operator $\Delta_f$ is positive semi-definite.
\el
\bp It is the general conclusion in the SQM. For $\a\in\mathcal{A}_{\C}^*$, set $\|\a\|^2=g(\a,\a)$, then
$$g(\a,\Delta_f a)=\|\bar{\p}_f\a\|^2+\|\bar{\p}_f^{\dag}\a\|^2.$$
Therefore, the spectra of $\Delta_f$ are non-negative.
\ep
Accordingly, $\zeta_2^i(s)$ is a holomorphic function for any $s\in\C$. It remains to consider $\zeta_1^i(s)$. We have to study the asymptotic expansion of heat trace via the heat kernel function.

The heat trace expansion as $t\rightarrow0+$ in our situation is quite different from the one on the compact manifolds. On the $n$-dimensional compact manifolds, there is an asymptotic expansion \cite{1992Heat} to the heat kernel function $p(x,y,t)$
$$p(x,y,t)\sim \frac{1}{(4\pi t)^{n/2}}e^{-\frac{d(x,y)^2}{4t}}\sum_{i=0}^{n}t^ib_i(x,y).$$
It follows that the heat trace admits an asymptotic expansion as a Laurent series of $t^{1/2}$ with the leading term of order $-\frac{n}{2}$. But here we are facing with the noncompact manifolds $\C^n$ and the unbounded potential, the trace expansion is more different. For example, assume that the base space is $\mathbb{R}^n$, let $L=-\sum_{i=1}^n\p_{x_i}^2+V(x)$ act on $L^2(\mathbb{R}^n)$, where the potential function $V$ is bounded from below, and $V\rightarrow+\infty$ as $|x|\rightarrow+\infty$. Then the local heat-kernel expansion in the diagonal (see in \cite{Cognola2006Heat}) can be partially summed over and
rewritten under the form
$$p(x,x,t)=\frac{1}{(4\pi t)^{n/2}}e^{-tV(x)}\sum_{i=0}^{\infty}U_{i}(x)t^{i},$$
where $U_{i}(x)$ are polynomials of derivatives of $V(x)$. And there may exist other fractional powers (not just the half integers), even logarithmic terms of $t$ in the heat trace asymptotic expansion as $t\rightarrow 0+$. For the potential $V=|x|^{Q}$, or $e^{|x|^{Q}}$, the result can be found in \cite{Cognola2006Heat}.

Therefore, it is much difficult to compute the asymptotic expansion for each $\Tr (e^{-t\Delta_f^k}-\Pi_k)$, $k=0,1,\ldots, 2n$. For simplicity, in the rest of the paper, let us focus on the non-degenerate homogenous polynomial $f_0$ of degree $p+1$ and its marginal or relevant deformation $f$, where $p\geq 1$.

First, if we choose the standard Hermtian metric $h=\frac{1}{2}\sum_{\nu=1}^ndz_{\nu}\otimes d\bar{z}_{\nu}$ on $\C^n$ and the standard basis $\{dz^I\wedge d\bar{z}^{J}\}_{I,J\subset\{1,\ldots,n\}|I|+|J|=k}$, then we can write $\Delta_f^k$ as a matrix operator $\Delta_f^k=-\Delta_{\bar{\p}}+B_k+|\nabla f|^2$Id, where $B_k$ represents the matrix of $L_f$. As we compute in Appendix A, the diagonal part ($w=z$) of heat kernel for $\Delta_f^k$ has the expansion of the form
\begin{enumerate}
  \item if $f$ is the marginal deformation, $p_{k}(z,z,t;u)=\frac{1}{(2\pi t)^n}e^{-2t\sum_{\nu}|\p_{z_{\nu}}f|^2}\sum_{a=0}^{\infty}U_a(z;u)t^a,$
  \item if $f$ is the relevant deformation, $p_{k}(z,z,t;u)=\frac{1}{(2\pi t)^n}e^{-2t\sum_{\nu}|\p_{z_{\nu}}f_0|^2}\sum_{a=0}^{\infty}\widetilde{U}_a(z;u)t^a,$
\end{enumerate}

where $U_a(z;u)$ (resp. $\widetilde{U}_a(z;u)$) can be solved by a certain recursion relation with $U_0(z;u)=1$ (resp. $\widetilde{U}_0(z;u)=1$) being given.

Since $\deg f_0=p+1$, we have
\begin{enumerate}
  \item for marginal deformation $f$, $\deg|\p f|^2=2p$ ;
  \item for relevant deformation $f$, $\deg|\p f_0|^2=2p$.
\end{enumerate}
Then for the two cases, we can do the rescaling $z\rightarrow y=t^{\frac{1}{2p}}z$, then the exponential part in $p(y,y,t;u)$ has no the factor $t$. Accordingly, we get the asymptotic expansion of $\Tr(e^{-t\Delta_f^k})$ as a Laurent series of $t^{\frac{1}{2p}}$ with the leading term of order $-\frac{(p+1)n}{p}$. 

Let us order the powers by $\a_1^k=-\frac{(p+1)n}{p}<\a_2^k<\cdots<\a_{i_0(k)}^k=0<\a_{i_0(k)+1}^k<\cdots$, then we can write
$$\Tr(e^{-t\Delta_f^k})=\sum_{1\leq a\leq i_0(k)}c_{a,k}(u,\bar{u})t^{\a_a^k}+\sum_{b>i_0(k)}c_{b,k}(u,\bar{u})t^{\a_b^k}.$$

\bpr\label{istr} Let $f_0(z)$ be a non-degenerate homogeneous polynomial on $\C^n$, and $f(z;u)=f_0(z)+\sum_{i=1}^{s}u^i\phi_i(z)$ be its relevant or marginal deformation. Then $\Tr(-1)^NN^i(e^{-t\Delta_f}-\Pi)$, $i\geq2$ has the following asymptotic expansion 
$$\Tr(-1)^NN^i(e^{-t\Delta_f}-\Pi)=\sum_{1\leq a\leq i_0}d_{a;i}(u,\bar{u})t^{\a_a}+\sum_{b>i_0}d_{b;i}(u,\bar{u})t^{\a_b}\quad \text{as }t\rightarrow0+,$$
where $\a_a,\a_b\in\mathbb{Q}$, $\a_1<\a_2<\cdots<\a_{i_0}=0<\a_{i_0+1}<\cdots$.
\epr

\bex Consider the deformation of $A_2$ singularity, $f=\frac{1}{3}z^3+uz$, we can use the construction method in Appendix A and the vanishing theorem to get: for $i\geq2$, 
\begin{flalign*}
&\Tr(-1)^NN^i(e^{-t\Delta_f}-\Pi)=(2^i-2)\Tr(e^{-t\Delta_f^2}-\Pi_2)=(2^i-2)\Tr(e^{-t\Delta_f^0}-\Pi_0)\\
=~&(2^i-2)\left[\frac{\sqrt{2\pi}}{8}t^{-\frac{3}{2}}-\frac{\sqrt{2\pi}}{8}|u|^2t^{-\frac{1}{2}}-\frac{1}{6}+O(t^{\frac{1}{2}})\right].
\end{flalign*}
\eex

Therefore, $\zeta_1^i(s)$ is holomorphic in the domain $\re(s)>\frac{(p+1)n}{p}$. Consider the regularized function as
$$\zeta_1^{R,i}(s)=\frac{1}{2\Gamma(s)}\int_0^1t^{s-1}\left[\Tr(-1)^NN^i(e^{-t\Delta_f}-\Pi)-\sum_{1\leq i\leq i_0}d_{a;i}(u,\bar{u})t^{\a_a}\right]dt$$
It is holomorphic in a neighborhood of $0\in\C$ and meromorphic for $\re(s)\geq0$. 
Or, $\zeta_1^i(s)$ extends to a meromorphic of function $$\zeta_1^{R,i}(s)=\frac{1}{2\Gamma(s)}\sum_{a\geq1}\frac{d_{a;i}(u,\bar{u})}{\a_a+s},\quad s\in\C,$$
which is holomorphic in a neighborhood of $0$.

\bd[Defintion 7.8 in \cite{Fan2016Torsion}] For $i\in\mathbb{N}$, the $i$-th torsion type invariants $T^i(f)$ is defined to be
\begin{equation}
\log T^i(f)=-(\zeta_1^{R,i}+\zeta_2^{i})'(0).
\end{equation}
\ed

\noindent\textbf{Remark:} Note that here $\log T^i(f)$ is a smooth function with respect to the parameters $u,\bar{u}$.
\vskip 0.1cm
Since we are interested in the 2nd zeta function, here we also pay our attention to the 2nd torsion type invariants. First, Proposition 7.9 in \cite{Fan2016Torsion} also holds in our case:
\bpr Let $(\C^{n_1},f_{01}(z))$ and $(\C^{n_2},f_{02}(w))$ be two non-degenerate homogeneous polynomials and $f_1(z;u)$ and $f_2(w;v)$ be their marginal or relevant deformations respectively, then we have the sum of the singularity $(\C^{n_1+n_2}, f_1(z;u)+f_2(w;v))$ and the identity of torsions
$$\log T^2(f_1\oplus f_2)=(-1)^{n_1}\mu(f_1)\log T^2(f_2)+(-1)^{n_2}\mu(f_2)\log T^2(f_1).$$
\epr
\bp Although the $f_1\oplus f_2$ may be no longer homogeneous, they are splittable. Analogously, we can construct $T^2(f_1\oplus f_2)$. Then proceed the proof of Proposition 7.9 in \cite{Fan2016Torsion}.
\ep

\subsection{Anomaly formula}

Recall that we have computed the transgression formula for $\Tr(-1)^NN^2(e^{-t\Delta_f}-\Pi)$ in Section 5. In this subsection, we apply it to the 2nd singularity torsion.

The main theorem of this section is:
\bt[Anomaly formula]\label{af} Let $f_0(z)$ be a non-degenerate and homogeneous polynomial on $\C^n$, and let $f(z;u)=f_0(z)+\sum_{i=1}^{s}u^i\phi_i(z)$ be its relevant or marginal deformation. Then
\begin{equation}\label{anomaly}
\bar{\p}_{\bar{j}}\p_i\log T^2(f)=(-1)^n\tr C_i\bar{C}_{\bar{j}}-\left(\Tr(-1)^N\int_0^t\phi_ie^{-t'\Delta_f}\bar{\phi}_{\bj}e^{-(t-t')\Delta_f}dt'\right)_1.
\end{equation}
In particular, when $f$ is the marginal deformation,
$$\left(\Tr(-1)^N\int_0^t\phi_ie^{-t'\Delta_f}\bar{\phi}_{\bj}e^{-(t-t')\Delta_f}dt'\right)_1=0.$$
\et

By the definition of the torsion type invariants, we have
\begin{equation}\label{logT2f}
\begin{aligned}
\log T^2(f)=~&-\frac{1}{2}\int_0^1\left(\Tr(-1)^NN^2(e^{-t\Delta_f}-\Pi)-\sum_{1\leq a\leq i_0}d_{a}(u,\bar{u})t^{\a_a}\right)\frac{dt}{t}\\
            &-\frac{1}{2}\int_1^{\infty}\Tr(-1)^NN^2(e^{-t\Delta_f}-\Pi)\frac{dt}{t}\\
            &-\frac{1}{2}\sum_{1\leq a<i_0}\frac{d_a(u,\bar{u})}{\a_a}+\frac{1}{2}~\Gamma'(1)d_{i_0}(u,\bar{u}).
\end{aligned}
\end{equation}

To prove the anomaly formula, we need the transgression formula \eqref{transgress2} and the following asymptotic analysis for the term
$$\Tr(-1)^N\phi_i\bar{\phi}_{\bar{j}}e^{-t\Delta_f}-\Tr(-1)^N\int_0^te^{-t'\Delta_f}\bar{\phi}_{\bar{j}}e^{-(t-t')\Delta_f}\Delta_f\phi_idt'.$$
More precisely, we have to analyze this term as $t\rightarrow\infty$ and its constant part.

\bl\label{inftyanalysis} As $t\rightarrow\infty$, we have
\begin{flalign*}
&\lim_{t\rightarrow\infty}\left(\Tr(-1)^N\phi_i\bar{\phi}_{\bar{j}}e^{-t\Delta_f}-\Tr(-1)^N\int_0^te^{-t'\Delta_f}\bar{\phi}_{\bar{j}}e^{-(t-t')\Delta_f}\Delta_f\phi_idt'\right)\\
=~&\tr(-1)^nC_i\bar{C}_{\bj}.
\end{flalign*}
\el
\bp Note that $\lim_{t\rightarrow\infty}e^{-t\Delta_f}=\Pi$, we have
\begin{flalign*}
&\lim_{t\rightarrow\infty}\left(\Tr(-1)^N\phi_i\bar{\phi}_{\bar{j}}e^{-t\Delta_f}-\Tr(-1)^N\int_0^te^{-t'\Delta_f}\bar{\phi}_{\bar{j}}e^{-(t-t')\Delta_f}\Delta_f\phi_idt'\right)\\
=~&\Tr(-1)^N\phi_i\bar{\phi}_{\bj}\Pi-\lim_{t\rightarrow\infty}\Tr(-1)^N\int_0^te^{-t'\Delta_f}\bar{\phi}_{\bar{j}}e^{-(t-t')\Delta_f}\Delta_f\phi_idt'\\
=~&\Tr(-1)^N\phi_i\bar{\phi}_{\bj}\Pi-\lim_{t\rightarrow\infty}\Tr(-1)^N\left(e^{-\frac{t}{2}\Delta_f}\int_0^{\frac{t}{2}}e^{-t'\Delta_f}\bar{\phi}_{\bar{j}}e^{-(\frac{t}{2}-t')\Delta_f}\Delta_fdt'\right.\\
 ~&+\left.\int_0^{\frac{t}{2}}e^{-t'\Delta_f}\bar{\phi}_{\bar{j}}e^{-(\frac{t}{2}-t')\Delta_f}\Delta_fdt'e^{-\frac{t}{2}\Delta_f}\right)\phi_i\\
=~&\Tr(-1)^N\phi_i\bar{\phi}_{\bj}\Pi-\Tr(-1)^N\Pi\lim_{t\rightarrow\infty}\int_0^{\frac{t}{2}}e^{-t'\Delta_f}\bar{\phi}_{\bar{j}}e^{-(\frac{t}{2}-t')\Delta_f}\Delta_fdt'\phi_i\\
 ~&-\Tr(-1)^N\lim_{t\rightarrow\infty}\int_0^{\frac{t}{2}}e^{-t'\Delta_f}\bar{\phi}_{\bar{j}}e^{-(\frac{t}{2}-t')\Delta_f}\Delta_fdt'\Pi\phi_i\\
\overset{(a)}=~&\Tr(-1)^N\phi_i\bar{\phi}_{\bj}\Pi-\Tr(-1)^N\Pi\lim_{t\rightarrow\infty}\int_0^{\frac{t}{2}}e^{-t'\Delta_f}\bar{\phi}_{\bar{j}}e^{-(\frac{t}{2}-t')\Delta_f}\Delta_fdt'\phi_i\\
=~&\Tr(-1)^N\phi_i\bar{\phi}_{\bar{j}}\Pi\\
~&-\Tr(-1)^N\Pi\lim_{t\rightarrow\infty}\int_0^{\frac{t}{2}}\left[\frac{d}{dt'}(e^{-t'\Delta_f}\bar{\phi}_{\bar{j}}e^{-(\frac{t}{2}-t')\Delta_f}dt')+\Delta_fe^{-t'\Delta_f}\bar{\phi}_{\bar{j}}e^{-(\frac{t}{2}-t')\Delta_f}dt'\right]\phi_i\\
\overset{(b)}=~&\Tr(-1)^N\phi_i\bar{\phi}_{\bar{j}}\Pi-\Tr(-1)^N\Pi\lim_{t\rightarrow\infty}(e^{-\frac{t}{2}\Delta_f}\bar{\phi}_{\bar{j}}-\bar{\phi}_{\bar{j}}e^{-\frac{t}{2}\Delta_f})\phi_i\\
=~&\Tr(-1)^N\Pi\bar{\phi}_{\bar{j}}\Pi\phi_i=\tr(-1)^nC_i\bar{C}_{\bj}
\end{flalign*}
In the equalities $(a)$ and $(b)$, we use the relations $\Delta_f\Pi=0$ and $\Pi\Delta_f=0$ respectively.
\ep

Next, we consider the constant part, i.e.
\begin{equation}\label{0term}
\left(\Tr(-1)^N\phi_i\bar{\phi}_{\bar{j}}e^{-t\Delta_f}-\Tr(-1)^N\int_0^te^{-t'\Delta_f}\bar{\phi}_{\bar{j}}e^{-(t-t')\Delta_f}\Delta_f\phi_idt'\right)_0,
\end{equation}
which is equivalent to
\begin{equation*}%\label{1term}
\left(\Tr(-1)^N\int_0^t\phi_ie^{-t'\Delta_f}\bar{\phi}_{\bj}e^{-(t-t')\Delta_f}dt'\right)_1.
\end{equation*}

To do this, we compute $\Tr(-1)^N\int_0^t\phi_ie^{-t'\Delta_f}\bar{\phi}_{\bj}e^{-(t-t')\Delta_f}dt'$ as follows:
\begin{flalign*}
  &\Tr(-1)^N\int_0^t\phi_ie^{-t'\Delta_f}\bar{\phi}_{\bj}e^{-(t-t')\Delta_f}dt'\\
=~&\Tr(-1)^N\int_0^t\phi_ie^{-t'\Delta_f}\bar{\phi}_{\bj}e^{t'\Delta_f}dt'e^{-t\Delta_f}\\
=~&\Tr(-1)^N\int_0^t\phi_i\left(\bar{\phi}_{\bj}-t'[\Delta_f, \bar{\phi}_{\bj}]+\frac{t'^2}{2!}[\Delta_f,[\Delta_f, \bar{\phi}_{\bj}]]-\cdots\right)dt'e^{-t\Delta_f}\\
=~&t\Tr(-1)^N\phi_i\bar{\phi}_{\bj}e^{-t\Delta_f}+\frac{t^2}{2!}\Tr(-1)^N\phi_i(2\bap_{\bar{\nu}}\bar{\phi}_{\bj}\p_{\nu})e^{-t\Delta_f}\\
 ~&+\frac{t^3}{3!}\Tr(-1)^N\phi_i\left(4\bar{\p}_{\bar{\mu}}\bar{\p}_{\bar{\nu}}\bar{\phi}_{\bj}\p_{\nu}\p_{\mu}+2\bar{\p}_{\bar{\nu}}\bar{\phi}_{\bj}\p_{\nu}L_f+4\bar{\p}_{\bar{\nu}}\bar{\phi}_{\bj}\p_{\nu}|\p f|^2\right)e^{-t\Delta_f}+\cdots
\end{flalign*}

For simplicity, we only consider the marginal deformation $f$ for the homogeneous polynomial $f_0$. Assume $\deg f=p+1$, we have
$$\deg |\p f|^2=\deg g=2p,\quad \deg\p_{\nu}\p_{\mu}f=p-1,\quad \deg(\det|\p^2f|^2)=2n(p-1).$$

\bl\label{constterm} Let $f_0$ be a non-degenerate homogeneous polynomial, and $f$ be its marginal deformation, then
$$\left(\Tr(-1)^N\int_0^t\phi_ie^{-t'\Delta_f}\bar{\phi}_{\bj}e^{-(t-t')\Delta_f}dt'\right)_1=0.$$
\el

The proof of this lemma is given via a scaling analysis and detailed computation as follows:

First consider the scaling $z\rightarrow t^{\frac{1}{2p}}z$, $w\rightarrow t^{\frac{1}{2p}}w$, using the recursion relation for $U_n$:
\begin{equation}\label{Unrecursion}
\begin{aligned}
U_n(z,w)=~&\frac{1}{r^n}\int_0^rs^{n-1}\left\{2\p_{\nu}\bap_{\bar{\nu}}U_{n-1}(z,w)-B(z)U_{n-1}(z,w)\right.\\
        ~&-2[\p_{\nu}g\bap_{\bar{\nu}}U_{n-2}(z,w)+\bap_{\bar{\nu}}g\p_{\nu}U_{n-2}(z,w)+\p_{\nu}\bap_{\bar{\nu}}gU_{n-2}(z,w)]\\
        ~&+\left.2\p_{\nu}g\bap_{\bar{\nu}}gU_{n-3}(z,w)\right\}ds,
\end{aligned}
\end{equation}
we know that
\begin{flalign}
&\left(\Tr(-1)^N\int_0^t\phi_ie^{-t'\Delta_f}\bar{\phi}_{\bj}e^{-(t-t')\Delta_f}dt'\right)_1\\
=~&\frac{1}{(2\pi)^n}\int_X\phi_i\bar{\phi}_{\bj}e^{-2|\p f|^2}\left(\str_{2n(p-1)-2} U_{2n+1}+\str_{2n(p-1)+2p-2} U_{2n+2}\right.\label{str2n+1}\\
~&\left.+\str_{2n(p-1)+4p-2} U_{2n+3}\right)dvol_z\label{str2n+3}\\
~&+\frac{1}{2!}\frac{1}{(2\pi)^n}\int_X\str\left(\phi_i(2\bar{\p}_{\bar{\nu}}\bar{\phi}_{\bj}\p_v)e^{-g(z,w)} U_{2n}(z,w)\big|_{w=z}\right)dvol_z\label{1t0}\\
~&+\frac{1}{3!}\frac{1}{(2\pi)^n}\int_X\phi_i(4\bar{\p}_{\bar{\nu}}\bar{\phi}_{\bj}\p_{\nu}|\p f|^2)e^{-2|\p f|^2}\str U_{2n}dvol_z\label{2t0}\\
~&+\frac{1}{3!}\frac{1}{(2\pi)^n}\int_X\phi_i(2\bar{\p}_{\bar{\nu}}\bar{\phi}_{\bj})e^{-2|\p f|^2}\str[(\p_{\nu}L_f)U_{2n-1}]dvol_z\label{3t0},
\end{flalign}
where $\str_{d}U_a$ denote the $\deg d$-component in $\str U_a$.

Next let us compute each term one by one. Recall that
$$\Delta_f=-2\sum\p_{\nu}\bap_{\bar{\nu}}+L_f+2|\p f|^2,$$
where
$$L_f=-2(\p_{\mu}\p_{\nu}f\iota_{\bar{\mu}}dz^{\nu}\wedge+\overline{\p_{\mu}\p_{\nu}f\iota_{\bar{\mu}}dz^{\nu}\wedge}).$$
\bl[\cite{Fan2016Torsion} Proposition 3.4] For $0\leq m\leq 2n$, we have
\begin{equation}\label{cliffordstr}
\str L_f^m: =\tr(-1)^NL_f^m=
\begin{cases}
0& 0\leq m<2n\\
(2n)!(-1)^n4^n|\det(\p^2f)|^2 & m=2n.
\end{cases}
\end{equation}
\el

Note that we use the matrix function $B_k$ to represent the action of $L_f$ on the space of $k$-forms. Then we have
\bc For $0\leq m\leq 2n$, we have
\begin{equation*}
\str B^m:=\sum_{k=0}^{2n}(-1)^k\tr B_k^m=
\begin{cases}
0& m\neq 2n\\
(2n)!(-1)^n4^n|\det(\p^2f)|^2 & m=2n.
\end{cases}
\end{equation*}
Furthermore, we have
$$\str U_m(z,z)=0, m<2n; \quad \str U_{2n}(z,z)=(-1)^n4^n|\det(\p^2f)|^2.$$
\ec

Using the recursion relation for $U_k$, we know that
\begin{itemize}
  \item $\str_{2n(p-1)-2}U_{2n+1}$ consists of $2n$ $(-B)$-terms and one $2\p_{\nu}\bap_{\bar{\nu}}$,
  \item $\str_{2n(p-1)+2p-2}U_{2n+2}$ consists of $2n$ $(-B)$-terms and one $-2(\p_{\nu}\bap_{\bar{\nu}}g+\p_{\nu}g\bap_{\bar{\nu}}+\bap_{\bar{\nu}}g\p_{\nu})$,
  \item $\str_{2n(p-1)+4p-2}U_{2n+3}$ consists of $2n$ $(-B)$-terms and one $2\p_{\nu}g\bap_{\bar{\nu}}g$.
\end{itemize}

To explicit to compute them, let us consider the iteration of $(-B)$-terms only. Set
\begin{equation}\label{Vnrecursion}
V_n(z,w)=\frac{1}{r^n}\int_0^rs^{n-1}[-B(z)V_{n-1}(z,w)]ds, \quad V_1(z,w)=\frac{1}{r}\int_0^r[-B(z)]ds.
\end{equation}
It is easy to see that we can expand $\tr V_n(z,w)$ as
\begin{flalign*}
\tr V_n(z,w)=~&(-1)^n\frac{1}{n!}\tr B(w)^n+a_n\tr\left\{B(w)^{n-1}[\p_{\nu}B(w)(z_{\nu}-w_{\nu})+\bap_{\bar{\nu}}B(w)(\bar{z}_{\bar{\nu}}-\bar{w}_{\bar{\nu}})]\right\}\\
             &+b_n\tr\left\{B(w)^{n-1}\p_{\nu}\bap_{\bar{\nu}}B(w)|z_{\nu}-w_{\nu}|^2\right\}+\cdots
\end{flalign*}
Note that $\p_z\bap_{\bar{z}}B=0$, we can compute that
$$a_1=-\frac{1}{2},\quad b_1=0.$$

Using the recursion relation for $V_n$, we get the recursion relation for the coefficients $a_n$, $b_n$ as follows.
$$a_{n+1}=-\frac{1}{n+2}\left[(-1)^n\frac{1}{n!}+a_n\right],\quad b_{n+1}=-\frac{1}{n+3}(2a_n+b_n), \quad n\geq1.$$
Moreover, by induction, we have
\bl For $n\geq2$,
$$a_n=(-1)^n\frac{1}{2}\times\frac{1}{(n-1)!}, \quad b_n=(-1)^n\frac{1}{4}\times\frac{1}{(n-2)!}.$$
\el

Now let us compute $\str_{2n(p-1)-2}U_{2n+1}$, $\str_{2n(p-1)+2p-2}U_{2n+2}$, $\str_{2n(p-1)+4p-2}U_{2n+3}$.

\noindent (1) $\str_{2n(p-1)-2}U_{2n+1}$:
\vskip 0.1cm
Using the recursion relation \eqref{Unrecursion} for $U_n$, we know that this term is computed in terms of the following iteration:
$$\underbrace{-B\leftarrow\cdots \leftarrow -B}_{2n-i}\leftarrow 2\p_{\nu}\bap_{\bar{\nu}}\leftarrow \underbrace{-B\leftarrow\cdots\leftarrow -B}_i,\quad i=2,\ldots, 2n,$$
where $\leftarrow$ denotes the iteration from \eqref{Unrecursion}. Then
\begin{flalign*}
\str_{2n(p-1)-2}U_{2n+1}=~&\sum_{i=2}^{2n}\frac{1}{(2n+1)\cdots(i+1)}\frac{1}{4}\times\frac{1}{(i-2)!}\str \left[B(z)^{2n-i}2\p_{\nu}\bap_{\bar{\nu}}B(z)^i\right]\\
                        =~&\frac{1}{4}\sum_{i=2}^{2n}\frac{i(i-1)}{(2n+1)!}\str \left[B(z)^{2n-1}2\p_{\nu}\bap_{\bar{\nu}}B(z)\right]\\
                        =~&\frac{1}{4}\frac{1}{(2n+1)!}\times\frac{1}{3}(2n-1)2n(2n+1)\str \left[B(z)^{2n-1}2\p_{\nu}\bap_{\bar{\nu}}B(z)\right]\\
                        =~&\frac{1}{12}2\p_{\nu}\bap_{\bar{\nu}}\frac{1}{(2n)!}\str B^{2n}.
\end{flalign*}

\noindent (2) $\str_{2n(p-1)+2p-2}U_{2n+2}$:
\vskip 0.1cm
Similarly, the iteration is given by
\begin{flalign*}
&\underbrace{-B\leftarrow\cdots \leftarrow -B}_{2n-i}\leftarrow-2\p_{\nu}\bap_{\bar{\nu}}g \leftarrow \underbrace{-B\leftarrow\cdots\leftarrow -B}_i,~i=0,\ldots,2n\\
&\underbrace{-B\leftarrow\cdots \leftarrow -B}_{2n-j} \leftarrow -2(\p_{\nu}g\bap_{\bar{\nu}}+\bap_{\bar{\nu}}g\p_{\nu})\leftarrow \underbrace{-B\leftarrow\cdots\leftarrow -B}_j, ~j=1,\ldots,2n
\end{flalign*}
Note that $g=2\int_0^1|\p f|^2(\tau(z-w)+w)d\tau$, then
$$\p_{\nu}g(z,w)\big|_{w=z}=\p_{\nu}|\p f|^2,~\bap_{\bar{\nu}}g(z,w)\big|_{w=z}=\bap_{\bar{\nu}}|\p f|^2,~ \p_{\nu}\bap_{\bar{\nu}}g(z,w)\big|_{w=z}=\frac{2}{3}\p_{\nu}\bap_{\bar{\nu}}|\p f|^2.$$
Therefore, we have
\begin{flalign*}
&\str_{2n(p-1)+2p-2}U_{2n+2}\\
=~&-2\sum_{i=0}^{2n}\frac{1}{(2n+2)\cdots(k+2)}\str\left[B^{2n-k}\frac{2}{3}\p_{\nu}\bap_{\bar{\nu}}|\p f|^2\frac{1}{k!}B^k\right]\\
&-2\sum_{j=1}^{2n}\frac{1}{(2n+2)\cdots(k+2)}\str\left[B^{2n-k}\p_{\nu}|\p f|^2\frac{1}{2}\times\frac{1}{(k-1)!}B^{k-1}\bap_{\bar{\nu}}B\right]\\
&-2\sum_{j=1}^{2n}\frac{1}{(2n+2)\cdots(k+2)}\str\left[B^{2n-k}\bap_{\bar{\nu}}|\p f|^2\frac{1}{2}\times\frac{1}{(k-1)!}B^{k-1}\p_{\nu}B\right]\\
=~&-\frac{1}{12}\frac{1}{(2n)!}\left(2\p_{\nu}\bap_{\bar{\nu}}|\p f|^2\str B^{2n}+2\p_{\nu}|\p f|^2\bap_{\bar{\nu}}\str B^{2n}+2\bap_{\bar{\nu}}|\p f|^2\p_{\nu}\str B^{2n}\right).
\end{flalign*}

\noindent (3) $\str_{2n(p-1)+4p-2}U_{2n+3}$:
\vskip 0.1cm
Similarly, the iteration is given by
$$\underbrace{-B\leftarrow\cdots \leftarrow -B}_{2n-i} \leftarrow 2\p_{\nu}g\bap_{\bar{\nu}}g\leftarrow \underbrace{-B\leftarrow\cdots\leftarrow -B}_i,\quad i=0,\ldots, 2n.$$
Then we have
\begin{flalign*}
&\str_{2n(p-1)+4p-2}U_{2n+3}\\
=~&2\sum_{i=0}^{2n}\frac{1}{(2n+3)\cdots(k+3)}\str\left[B^{2n-i}\p_{\nu}|\p f|^2\bap_{\bar{\nu}}|\p f|^2\frac{1}{k!}B^k\right]\\
                           =~&\frac{1}{12}\frac{1}{(2n)!}2\p_{\nu}|\p f|^22\bap_{\bar{\nu}}|\p f|^2\str B^{2n}.
\end{flalign*}

Thus, we obtain
\begin{flalign*}
&e^{-2|\p f|^2}\left(\str_{2n(p-1)-2}U_{2n+1}+\str_{2n(p-1)+2p-2}U_{2n+2}+\str_{2n(p-1)+4p-2}U_{2n+3}\right)\\
=~&\frac{1}{12}\frac{1}{(2n)!}\p_{\nu}\bap_{\bar{\nu}}\left(e^{-2|\p f|^2}\str B^{2n}\right).
\end{flalign*}
After a rescaling, we see that
\begin{flalign*}
\eqref{str2n+1}+\eqref{str2n+3}=~&\frac{1}{12}\int_X\phi_i\bar{\phi}_{\bar{j}}\p_{\nu}\bap_{\bar{\nu}}\left(e^{-|\p f|^2}(-1)^n|\det|\p^2f||^2\right)dvol_z\\
           =~&\frac{1}{12}\int_X(\p_{\nu}\phi_i\bap_{\bar{\nu}}\bar{\phi}_{\bar{j}})e^{-|\p f|^2}(-1)^n|\det(\p^2f)|^2dvol_z.
\end{flalign*}

Next, let us compute the last three terms.\\

\noindent (4)  $\str\left(\phi_i(2\bar{\p}_{\bar{\nu}}\bar{\phi}_{\bj}\p_z)e^{-g(z,w)}U_{2n}(z,w)\big|_{w=z}\right)$:
\begin{flalign*}
&\str\left(\phi_i(2\bar{\p}_{\bar{\nu}}\bar{\phi}_{\bj}\p_{\nu})e^{-g(z,w)}U_{2n}(z,w)\big|_{w=z}\right)\\
=~&2\phi_i\bar{\p}_{\bar{\nu}}\bar{\phi}_{\bj}\str\left[\p_{\nu}(e^{-g(z,w)}U_{2n}(z,w))\big|_{w=z}\right]\\
=~&2\phi_i\bar{\p}_{\bar{\nu}}\bar{\phi}_{\bj}e^{-2|\p f|^2}\str\left[-\p_{\nu}|\p f|^2\frac{1}{(2n)!}B^{2n}+\frac{1}{2}\times\frac{1}{(2n-1)!}B^{2n-1}\p_{\nu}B\right]\\
=~&\frac{1}{2}\times2\phi_i\bar{\p}_{\bar{\nu}}\bar{\phi}_{\bj}\p_{\nu}\left(e^{-2|\p f|^2}\frac{1}{(2n)!}\str B^{2n}\right).
\end{flalign*}

Then, after a rescaling, we have
\begin{flalign*}
\eqref{1t0}=~&\int_X 2\phi\bap_{\bar{\nu}}\bar{\phi}_{\bj}\p_{\nu}\left(e^{-2|\p f|^2}\frac{1}{(2n)!}\str B^{2n}\right)\\
           =~&-\frac{1}{4}\int_X\p_{\nu}\phi_i\bap_{\bar{\nu}}\bar{\phi}_{\bj}e^{-|\p f|^2}(-1)^n|\det(\p^2f)|^2dvol_z.
\end{flalign*}

Similarly, we can compute $\eqref{2t0}$, $\eqref{3t0}$ as follow.
\vskip 0.1cm

\noindent (5) $\eqref{2t0}$:
\begin{flalign*}
\eqref{2t0}=~&\frac{1}{3!}\int_X\phi_i(\bar{\p}_{\bar{\nu}}\bar{\phi}_{\bj})(\p_{\nu}|\p f|^2)e^{-|\p f|^2}(-1)^n|\det(\p^2f)|^2dvol_z\\
          =~&\frac{1}{3!}\int_X\p_{\nu}\phi_i(\bar{\p}_{\bar{\nu}}\bar{\phi}_{\bj})e^{-|\p f|^2}(-1)^n|\det(\p^2f)|^2dvol_z\\
           ~&+\frac{1}{3!}\int_X\phi_i(\bar{\p}_{\bar{\nu}}\bar{\phi}_{\bj})e^{-|\p f|^2}(-1)^n\p_{\nu}|\det(\p^2f)|^2dvol_z.
\end{flalign*}

\noindent (6) $\eqref{3t0}$:
\begin{flalign*}
\eqref{3t0}=~&\frac{1}{3!}\int_X\phi_i(\bar{\p}_{\bar{\nu}}\bar{\phi}_{\bj})e^{-2|\p f|^2}(-1)^{2n-1}\frac{1}{(2n-1)!}\str (B^{2n-1}\p_{\nu}B)\\
           =~&-\frac{1}{3!}\int_X\phi_i(\bar{\p}_{\bar{\nu}}\bar{\phi}_{\bj})e^{-|\p f|^2}(-1)^n\p_{\nu}|\det(\p^2f)|^2dvol_z.
\end{flalign*}

In summary, we have
\begin{flalign*}
&\left(\Tr(-1)^N\int_0^t\phi_ie^{-t'\Delta_f}\bar{\phi}_{\bj}e^{-(t-t')\Delta_f}dt'\right)_1\\
=~&\left(\frac{1}{12}-\frac{1}{4}+\frac{1}{6}\right)\int_X\p_{\nu}\phi_i(\bar{\p}_{\bar{\nu}}\bar{\phi}_{\bj})e^{-|\p f|^2}(-1)^n|\det(\p^2f)|^2dvol_z\\
=~&0.
\end{flalign*}

\bp[Proof of Theorem \ref{af}] We learn from the transgression formula that $\bar{\p}_{\bj}\p_id_{i_0;2}=0.$\footnote{In fact, we can use the asymptotic analysis above to obtain $$d_{i_0;2}=0\quad \text{for } n\geq2.$$}
Using the equation \eqref{logT2f}, Theorem \ref{transgress2}, and Lemmas \ref{inftyanalysis}, \ref{constterm}, we complete the proof.
\ep

\noindent\textbf{Remarks:} (1) It is obvious to see that the two terms in r.h.s of anomaly formula \eqref{anomaly} can give us two K\"{a}hler metrics based on the deformation parameter space respectively. Here $\tr C_i\bar{C}_{\bar{j}}$ is K\"{a}hler because of the $tt^*$ equations:
$$\p_k\tr C_i\bar{C}_{\bar{j}}=\tr((D_kC_i)\bar{C}_{\bar{j}})=\tr((D_iC_k)\bar{C}_{\bar{j}})=\p_i\tr C_k\bar{C}_{\bar{j}}.$$
Meanwhile, note that
$$\tr(-1)^nC_i\bar{C}_{\bar{j}}=\lim_{L\rightarrow\infty}\Tr[(-1)^N\phi_ie^{-L\Delta_f}\bar{\phi}_{\bar{j}}e^{-L\Delta_f}].$$
If we look at it in a 2d TFT version, it is a torus with the fields $\phi_i$ and $\bar{\phi}_{\bar{j}}$ inserted on the left and right side of a flat torus respectively which are infinitely separated by two long tubes each with perimeter 1. Then it gives us the genus 1 information.
\vskip 0.1cm
\noindent(2) The holomorphic anomaly equation \cite{Bershadsky1994Kodaira} of the partition function $F_1$ for the Calabi-Yau 3-fold is given by
$$\bar{\p}_{\bar{j}}\p_iF_1=\frac{1}{2}\tr C_i\bar{C}_{\bar{j}}-\frac{\Tr(-1)^F}{24}G_{i\bar{j}},$$
where $G_{i\bar{j}}$ the Weil-Peterson metric which is determined by the special geometry relation (genus 0 information).
\vskip 0.1cm
\noindent(3) In \cite{Coates2015A}, Coates and Iritani also derived the anomaly equations for the genus $g$ partition function which they denoted by $C^{(g)}$. They started from the geometric and Givental quantizations, and
pointed out that the anomaly equations correspond to the choices of polarization for the symplectic space. The complex conjugate polarization gives the holomorphic anomaly equations. Their result for $C^{(1)}$ is
$$\bar{\p}_{\bj}\p_iC^{(1)}=-\frac{1}{2}\tr C_i\bar{C}_{\bar{j}}.$$
It is worth to mention that they introduced the descendants in their work. \\

\section*{Acknowledgements}
I am grateful to my advisor Professor Huijun Fan who introduces me the heat kernel theory associated to the singularity. I would also like to thank Professor Si Li and Professor Xiaojun Chen for giving me some useful advices. \\

\appendix

\begin{appendices}
\section{The proof of Theorems \ref{kernel} and \ref{deformedkernel}}

\vskip 0.1cm
In general, the basic idea to the construction of the heat kernel function is:
\begin{enumerate}
  \item constructing an approximate solution $p_K(z,w,t)$, and studying the remainder $r_K(z,w,t)=(\p_t+\Delta_{f_0})p_K(z,w,t)$,
  \item proving the convergence of the series $\sum_{i=0}^{\infty}p_K^i$ to the heat kernel function $p(z,w,t)$, where $p_K^i$'s are defined through the convolution, that is, $p_K^i=p_K*\underbrace{r_K*\cdots *r_K}_{i}$.
\end{enumerate}
The step (2) is standard in \cite{1992Heat}, the key point in our proof is (1). Now let us rewrite the proof of Theorem 4.1. For simplicity, we choose the metric $h=\frac{1}{2}\sum_{\nu=1}^{n}dz_{\nu}\otimes d\bar{z}_{\nu}$ on $\C^n$.
\vskip 0.1cm
Note that when $\Delta_{f_0}$ acts on each $k$-forms, we can choose a basis for the space of $k$-forms, and write $\Delta_{f_0}$ as a matrix-valued differential operators $\Delta_{f_0}^k$
$$\Delta_{f_0}^k=-2\sum_{\nu=1}^{n}\p_{z_{\nu}}\p_{\bar{z}_{\nu}}+B_{k}+2|\p f_0|^{2},$$
where the non-zero entry of the matrix $B_k$ is either $\p_{\mu}\p_{\nu}f_0$ or $\overline{\p_{\mu}\p_{\nu}f_0}$.

When $\Delta_{f_0}$ acts on the 0-forms and $2n$-forms, we have
$$\Delta_{f_0}^{0}=\Delta_{f_0}^{2n}=-2\sum_{\nu=1}^{n}\p_{z_{\nu}}\p_{\bar{z}_{\nu}}+2|\p f_0|^{2}.$$

Let us consider the heat kernel function of this case first.\\
Define
$$\mathcal{E}_{0}=\frac{1}{(2\pi t)^{n}}\exp{\left(-\frac{|z-w|^2}{2t}\right)},\quad \m{E}_{1}=\exp{(-tg)},$$\\
where $g=g(z,w)=2\int_{0}^{1}\sum_{\nu=1}^{n}\left|\frac{\p f_0}{\p z_{\nu}}\right|^{2}(\tau(z-w)+w)d\tau$.
Then it is easy to obtain
$$(z_{\nu}-w_{\nu})\p_{\nu}g+(\bar{z}_{\nu}-\bar{w}_{\nu})\p_{\bar{\nu}}g+g=V:=2|\p f_0|^2.$$

Now fix a sufficiently large $K\in\mathbb{N}$, we consider an approximation solution $p_{K}(z,w,t)$ of heat kernel equation for small $t$ in the sense that
$$r_{K}(z,w,t)=\frac{\p p_{K}(z,w,t)}{\p t}+\Delta_{f_0}^{0}p_{K}(z,w,t)=O(t^{\alpha}) \quad\text{ for some } \alpha>0.$$
Let us set $p_K(z,w,t)$ in the form
$$p_{K}(z,w,t)=\m{E}_{0}\m{E}_{1}\sum_{a=0}^{K}U_{a}(z,w)t^{a}.$$
Let $U=\sum_{a=0}^{K}U_{a}(z,w)t^{a}$, then
\begin{flalign*}
(\p_{t}+\Delta_{f_0}^{0})p_{K}(z,w,t)= ~&(\p_{t}-2\p_{z_{\nu}}\p_{\bar{z}_{\nu}}+V)\m{E}_{0}\m{E}_{1}U\\
                             =~&\m{E}_{0}\m{E}_{1}\left\{\sum_{a=0}^{K}(r\p_{r}U_{a}+aU_{a})t^{a-1}-2\sum_{a=0}^{K}\p_{z_{\nu}}\p_{\bar{z}_{\nu}}U_{a}t^{a}+2\sum_{a=0}^{K}\right.(g_{z_{\nu}\bar{z}_{\nu}}U_{a}\\
                               ~&+g_{z_{\nu}}\p_{\bar{z}_{\nu}}U_{a}+g_{\bar{z}_{\nu}}\p_{z_{\nu}}U_{a})t^{a+1}-\left.2g_{z_{\nu}}g_{\bar{z}_{\nu}}\sum_{a=0}^{K}U_{a}t^{a+2}\right\}
\end{flalign*}
Where $r=|z-w|$ and we omit $\sum_{\nu=1}^{n}$.\\

\noindent Now look at the coefficient of each $t^{a-1}$ in the brace for $a\geq 0$, then we get
\begin{equation*}
\begin{aligned}
&\text{Coeff}(t^{-1})=r\p_{r}U_{0}\\
&\text{Coeff}(t^0)=r\p_{r}U_{1}+U_{1}-2\p_{z_{\nu}}\p_{\bar{z}_{\nu}}U_{0}\\
&\text{Coeff}(t^1)=r\p_{r}U_{2}+2U_{2}-2\p_{z_{\nu}}\p_{\bar{z}_{\nu}}U_{1}+2(g_{z_{\nu}\bar{z}_{\nu}}U_{0}+g_{z_{\nu}}\p_{\bar{z}_{\nu}}U_{0}+g_{\bar{z}_{\nu}}\p_{z_{\nu}}U_{0})\\
&\text{Coeff}(t^a)=r\p_{r}U_{a+1}+(a+1)U_{a+1}-2\p_{z_{\nu}}\p_{\bar{z}_{\nu}}U_{a}+2(g_{z_{\nu}\bar{z}_{\nu}}U_{a-1}+g_{z_{\nu}}\p_{\bar{z}_{\nu}}U_{a-1}\\
&\quad\quad\quad\quad+g_{\bar{z}_{\nu}}\p_{z_{\nu}}U_{a-1})-2g_{z_{\nu}}g_{\bar{z}_{\nu}}U_{a-2},\quad a\geq2.
\end{aligned}
\end{equation*}
Consider the equations $\text{Coeff}(t^{a-1})=0$ for $a\geq 0$, then we can solve each $U_{a}$, $j=0,...,K$.
The first equation is
$$r\p_{r}U_{0}=0.$$
Then we can set $U_{0}=1$(normalization), and the second equation becomes
$$r\p_{r}U_{1}+U_{1}=0.$$
Then we can solve that $U_{1}=0$. And the third equation becomes
$$r\p_{r}U_{2}+2U_{2}+2g_{z_{\nu}\bar{z}_{\nu}}=0.$$
Then we can solve that $$U_{2}=\frac{1}{r^2}\int_{0}^{r}-2sg_{z\bar{z}}(z,w)ds.$$
For $a\geq 3$,
$$U_{a}=\frac{1}{r^{a}}\int_{0}^{r}2s^{a-1}\left(\p_{z_{\nu}}\p_{\bar{z}_{\nu}}U_{a-1}-g_{z_{\nu}\bar{z}_{\nu}}U_{a-2}-g_{z_{\nu}}\p_{\bar{z}_{\nu}}U_{a-2}-g_{\bar{z}_{\nu}}\p_{z_{\nu}}U_{a-2}+g_{z_{\nu}}g_{\bar{z}_{\nu}}U_{a-3}\right)ds.$$

Moreover, by induction, (see in \cite{Fan2016Torsion}) $U_{a}$ has the form
$$U_{a}(z,w)=U_{a}(z-w,w).$$
Then we can write the formula of $U_{a}$ as
\begin{equation}\label{eq:U}
\begin{aligned}
  &U_{a}(z,w)=U_{a}(z-w,w)\\
=~&\int_{0}^{1}\left\{\p_{z_{\nu}}\p_{\bar{z}_{\nu}}U_{a-1}(\tau(z-w),w)-g_{z_{\nu}\bar{z}_{\nu}}(\tau(z-w),w)U_{a-2}(\tau(z-w),w)\right.\\
 &-g_{\bar{z}_{\nu}}(\tau(z-w),w)\p_{z_{\nu}}U_{a-2}(\tau(z-w),w)-g_{z_{\nu}}(\tau(z-w),w)\p_{\bar{z}_{\nu}}U_{a-2}(\tau(z-w),w)\\
 &\left.+g_{z_{\nu}}g_{\bar{z}_{\nu}}(\tau(z-w),w)U_{a-3}(\tau(z-w),w)\right\}\tau^{a-1}d\tau.
\end{aligned}
\end{equation}

Now we see that $U_{a}(z,w)$ is uniquely determined if we fix $U_{0}=1$, then get the remainder
\begin{flalign*}
r_{K}(z,w,t)=~&(\p_{t}+\Delta_{f}^{0})p_{K}(z,w,t)\\
              =~&2\m{E}_{0}\m{E}_{1}\left[(-\p_{z_{\nu}}\p_{\bar{z}_{\nu}}U_{K}+g_{z_{\nu}\bar{z}_{\nu}}U_{K-1}+g_{z_{\nu}}\p_{\bar{z}_{\nu}}U_{K-1}+g_{\bar{z}_{\nu}}\p_{z_{\nu}}U_{K-1}-g_{z_{\nu}}g_{\bar{z}_{\nu}}U_{K-2})t^{K}\right.\\
                &\left.+(g_{z_{\nu}\bar{z}_{\nu}}U_{K}+g_{z_{\nu}}\p_{\bar{z}_{\nu}}U_{K}+g_{\bar{z}_{\nu}}\p_{z_{\nu}}U_{K}-g_{z_{\nu}}g_{\bar{z}_{\nu}}U_{K-1})t^{K+1}-g_{z_{\nu}}g_{\bar{z}_{\nu}}U_{K}t^{K+2}\right]\\
              =~:&~\m{E}_{0}\m{E}_{1}\widetilde{U}.
\end{flalign*}

For the approximation solution $p_{K}(z,w,t)$ and the remainder $r_{K}(z,w,t)$, we have the following result.

\bpr
\label{estimater}
Let $l\in\mathbb{N}$, denote $dvol_w=\left(\frac{i}{2}\right)^ndw_1\wedge d\bar{w}_1\wedge\cdots\wedge dw_n\wedge d\bar{w}_n$,
\begin{enumerate}
  \item For $\forall~ T>0$, $p_{K}(z,w,t)$, $0 \leq t\leq T$, define a uniformly bounded family of operators $P_{K}^{t}$ on $\mathcal{S}(\mathbb{C}^{n})$ and %(the space of $l$-differential functions with compact support)
$$\lim_{t\rightarrow 0}\parallel P_{K}^{t}s-s\parallel_{l}=0.$$
  \item Set $\delta=\frac{1-3(q_{1}-q_{n})}{1-q_1}$. When $\delta>0~(i.e.~q_{1}-q_{n}<\frac{1}{3})$ the remainder $r_{K}(z,w,t)=(\p_{t}+\Delta_{f}^{0})p_{K}(z,w,t)$ satisfies the estimate
$$|r_{K}(z,w,t)|\leq \m{E}_{0}t^{\frac{K}{3}\delta-\frac{2}{3(1-q_1)}}H(t),$$
and moreover
$$\|r_{K}(z,w,t)\|\leq t^{\frac{K}{3}\delta-\frac{2}{3(1-q_1)}-n}H(t),\quad \int_{\mathbb{C}^n}|r_{K}(z,w,t)|dw\wedge d\bar{w}\leq t^{\frac{K}{3}\delta-\frac{2}{3(1-q_1)}}H(t),$$
where
$H(t)$ is a polynomial of fractional powers of $t$ and it depends on $f_0$ and $K$.
\end{enumerate}
\epr

\bp Fix $K$, for $\forall s\in \mathcal{S}(\mathbb{C}^n)$, we have
$$P_{K}^{t}s(z,\bar{z})=\frac{1}{(2\pi t)^{n}}\int_{\mathbb{C}^n}e^{-\frac{|z-w|^2}{2t}}e^{-tg(z,w)}\sum_{a=0}^{K}U_{a}(z,w)t^{a}s(w,\bar{w})dvol_w.$$
Set $w-z=\sqrt{t}v$, then the integral becomes
$$\frac{1}{(2\pi)^{n}}\int_{\mathbb{C}}e^{-|v|^2}e^{-tg(z,z+\sqrt{t}v)}\sum_{a=0}^{K}U_{a}(z,z+\sqrt{t}v)t^{a}s(z+\sqrt{t}v,\bar{z}+\sqrt{t}\bar{v})dvol_v.$$
Note that we have set $U_{0}=1$, from which (1) follows. Such an operator $P_{K}^{t}$ is called parametrix for the heat equation.
\ep

The proof of (2) depends on the following lemma:

\bl[\bf{Dimension Argument}] Denote by $p_g^a$ the powers of $g$ and $p_{\p}^a$ the powers of derivatives in each term of $U_a(z,w)$ for $a\geq2$, then
$$2p_g^a+p_{\p}^a=2a,\quad p_{\p}^a>0 \quad \text{and}\quad p_g^a\leq p_{\p}^a.$$
\el
\bp
This can be shown by induction. First, when $a=2$, we have $p_g^2=1, p_{\p}^2=2$. Assume that for any $2\leq l\leq a-1$, the relations hold. Then for $U_{a}$, the relation follows from the formula \eqref{eq:U}.
\ep

An immediate corollary is
\bc For each $U_a(z,w)$, $p_g^a\leq [\frac{2a}{3}]$.
\ec

For simplicity, we introduce some notations:
\begin{itemize}
  \item $g_{\mu}=2\int_{0}^{1}|\p_{\mu}f_0|^{2}(\tau(z-w)+w)d\tau$,
  \vskip 0.2cm
  \item Let $(\p_{z_1}^{l_1}\p_{\bar{z}_1}^{l_1}...\p_{z_n}^{l_n}\p_{\bar{z}_n}^{l_n}, g_{1}^{m_1}...g_{n}^{m_n})$ be a unified representation for the monomial of $z-w,w,\bar{z}-\bar{w},\bar{w}$ in
      $$\p_{z_1}^{l_1}\p_{\bar{z}_1}^{l_1}...\p_{z_n}^{l_n}\p_{\bar{z}_n}^{l_n}[g_{1}^{m_1}...g_{n}^{m_n}].$$
\end{itemize}

Then each monomial of $z-w,w,\bar{z}-\bar{w},\bar{w}$ in $U_a$ can be found in $$\{(\p_{z_1}^{l_1}\p_{\bar{z}_1}^{l_1}...\p_{z_n}^{l_n}\p_{\bar{z}_n}^{l_n}, g_{1}^{m_1}...g_{n}^{m_n})~\big|~l_{1}+...+l_{n}=l=p_{\p}^a/2, m_{1}+...+m_{n}=m=p_g^a, l+m=a, 2l\geq m\}.$$

\noindent\textbf{Remark}: Since we choose the standard metric on $\C^{n}$, $\p_{z_{\mu}}$ and $\p_{\bar{z}_{\mu}}$ come in pairs. Of course, the existence of the heat kernel holds for constant positive-definite metrics.\\

\bp[Proof of Proposition \ref{estimater} (2)]
Look at the remainder
\begin{flalign*}
r_{K}(z,w,t)=~&2\m{E}_{0}\m{E}_{1}\left[(-\p_{z_{\nu}}\p_{\bar{z}_{\nu}}U_{K}+g_{z_{\nu}\bar{z}_{\nu}}U_{K-1}+g_{z_{\nu}}\p_{\bar{z}_{\nu}}U_{K-1}+g_{\bar{z}_{\nu}}\p_{z_{\nu}}U_{K-1}-g_{z_{\nu}}g_{\bar{z}_{\nu}}U_{K-2})t^{K}\right.\\
                &\left.+(g_{z_{\nu}\bar{z}_{\nu}}U_{K}+g_{z_{\nu}}\p_{\bar{z}_{\nu}}U_{K}+g_{\bar{z}_{\nu}}\p_{z_{\nu}}U_{K}-g_{z_{\nu}}g_{\bar{z}_{\nu}}U_{K-1})t^{K+1}-g_{z_{\nu}}g_{\bar{z}_{\nu}}U_{K}t^{K+2}\right]\\
              =~:&~\m{E}_{0}\m{E}_{1}\widetilde{U}.
\end{flalign*}
The term in coefficient($t^a$) ($a=K,K+1,K+2$) on the right hand side is contained in
$$\text{Span}_{\C}\{(\p_{z_1}^{l_1}\p_{\bar{z}_1}^{l_1}...\p_{z_n}^{l_n}\p_{\bar{z}_n}^{l_n}, g_{1}^{m_1}...g_{n}^{m_n})~\big|~l+m=a+1, 2l\geq m \}.$$

Without loss of generality, assume $q_1\geq q_2\geq\ldots\geq q_n$, i.e. $q_M=q_1, q_m=q_n$, such that $f_0$ has weight 1,  then $\w(g_{\nu})=\w( |\p_{\nu}f_0|^{2})=2-2q_{\nu}$, and $2-2q_{1}\leq 2-2q_{2}\leq...\leq 2-2q_{n}$ by assumption. We do the rescailing for each $z_{\nu}$
$$z_{\nu}~\longmapsto~ t^{\frac{q_{\nu}}{2-2q_1}}z_{\nu}, \text{ for } \nu=1,...,n.$$
Then for $0<t\leq 1$
\begin{flalign*}
\exp(-tg)~&\longmapsto~ \exp(-\sum_{\nu=1}^{n}t^{\frac{q_{\nu}-q_{1}}{1-q_{1}}}g_{\nu})\leq \exp(-\sum_{\nu=1}^{n}g_{\nu}),\\
(\p_{z_1}^{l_1}\p_{\bar{z}_1}^{l_1}...\p_{z_n}^{l_n}\p_{\bar{z}_n}^{l_n}, g_{1}^{m_1}...g_{n}^{m_n})t^{a}~&\longmapsto~
(\p_{z_1}^{l_1}\p_{\bar{z}_1}^{l_1}...\p_{z_n}^{l_n}\p_{\bar{z}_n}^{l_n}, g_{1}^{m_1}...g_{n}^{m_n})t^{a-\overrightarrow{m}\cdot\frac{1-\overrightarrow{q}}{1-q_1}+2\overrightarrow{l}\cdot\frac{\overrightarrow{q}}{2-2q_1}},
\end{flalign*}
where $\overrightarrow{l}=(l_{1},...,l_{n}), \overrightarrow{m}=(m_{1},...,m_{n}), \overrightarrow{q}=(q_{1},...,q_{n})$.
In the remainder $r_{K}(z,w,t)$, we have $a=K,a=K+1,a=K+2$, and
\begin{flalign*}
 &\min_{\overrightarrow{l},\overrightarrow{m}}\left(a-\overrightarrow{m}\cdot\frac{1-\overrightarrow{q}}{1-q_1}+\overrightarrow{l}\cdot\frac{\overrightarrow{q}}{1-q_1}\right)\\
=~&a-[\frac{2(a+1)}{3}]\times\frac{1-q_n}{1-q_1}+(a-[\frac{2(a+1)}{3}])\times \frac{q_n}{1-q_{1}}\\
\geq~ & \frac{a}{3}\frac{1-3(q_{1}-q_{n})}{1-q_1}-\frac{2}{3(1-q_1)}:= \frac{a}{3}\delta-\frac{2}{3(1-q_1)}.
\end{flalign*}

Since $e^{-g}(\p_{z_1}^{l_1}\p_{\bar{z}_1}^{l_1}...\p_{z_n}^{l_n}\p_{\bar{z}_n}^{l_n}, g_{1}^{m_1}...g_{n}^{m_n})\leq C$, the constant depends on $l,m$.
Therefore, we obtain
$$|r_{K}(z,w,t)|\leq \m{E}_{0}(t^{ \frac{K}{3}\delta-\frac{2}{3(1-q_1)}}H_{1}(t)+t^{\frac{K+1}{3}\delta-\frac{2}{3(1-q_1)}}H_{2}(t)+t^{ \frac{K+2}{3}\delta-\frac{2}{3(1-q_1)}}H_{3}(t)):=\m{E}_0t^{ \frac{K}{3}\delta-\frac{2}{3(1-q_1)}}H(t),$$
where $H(t)$ depends on $f_0$ and $K$. When $q_{1}-q_{n}<\frac{1}{3}$, we have $\delta>0$.
\ep
\vskip 0.1cm

\noindent\textbf{Remark}: Here we choose $0<t\leq 1$. Actually, for $t>1$, we can do another rescaling
$$z_{\nu}~\longmapsto~ t^{\frac{q_{\nu}}{2-2q_n}}z_{\nu}.$$
\vskip 0.2cm

There are more estimates about the derivatives of $r_{K}(z,w,t)$. First we have
\begin{flalign*}
\p_{z_{\nu}}r_{K}(z,w,t)&=(\p_{z_{\nu}}\m{E}_{0})\m{E}_{1}\widetilde{U}+\m{E}_{0}(\p_{z_{\nu}}\m{E}_{1})\widetilde{U}+\m{E}_{0}\m{E}_{1}\p_{z_{\nu}}\widetilde{U}\\
                &=\m{E}_{0}\m{E}_{1}\left(-\frac{\bar{z}_{\nu}-\bar{w}_{\nu}}{2t}\widetilde{U}-h_{z_{\nu}}\widetilde{U}t+\p_{z_{\nu}}\widetilde{U}\right).
\end{flalign*}
After the first rescaling, the lowest power of $t$ on the right hand side becomes
\begin{flalign*}
  &\min\{\frac{K}{3}\delta-\frac{2}{3(1-q_1)}-1-\frac{q_{\nu}}{2-2q_1}, \frac{K}{3}\delta-\frac{2}{3(1-q_1)}+1-\frac{1-q_{\nu}}{1-q_1}+\frac{q_{\nu}}{2-2q_1}\}\\
=~& \frac{K}{3}\delta-\frac{2}{3(1-q_1)}-1-\frac{q_{\nu}}{2-2q_1},
\end{flalign*}
since $q_{1}-q_{\nu}<\frac{1}{2}$.
Then we obtain
$$|\p_{z_{\nu}}r_{K}(z,w,t)|<\m{E}_{0}t^{\frac{K\delta}{3}-\frac{2}{3(1-q_1)}-1-\frac{q_{\nu}}{2-2q_1}}H(t).$$
Where $H(t)$ is different from the one before, but still depends on $f_0$ and $K$.\\
For $l\in\mathbb{N}$ and $\frac{K\delta}{3}-\frac{2}{3(1-q_1)}-n-l_{0}-\frac{l_{0}q_1}{2-2q_1}>0$, we have the estimate
\begin{flalign*}
\|r_{K}(z,w,t)\|_{l_0}\leq ~& t^{\frac{K\delta}{3}-\frac{2}{3(1-q_1)}-1-n-l_{0}-\frac{l_{0}q_1}{2-2q_1}}H_{l_{0},K}(t),\\
\p_{z}^{\overrightarrow{l}_{0}'}\p_{\bar{z}}^{\overrightarrow{l}_{0}-\overrightarrow{l}_{0}'}\int_{\mathbb{C}^n}|r_{K}(z,w,t)||dvol_w| \leq ~& t^{\frac{K\delta}{3}-\frac{2}{3(1-q_1)}-l-\frac{lq_1}{2-2q_1}}H_{l_{0},K}(t).
\end{flalign*}

Now we consider the operator $P_{K}^{i}$ defined by the kernel
$$p_{K}^{i}(z,w,t)=\int_{t\Delta_{i}}dt_{1}dt_{2}...dt_{i}\int_{\mathbb{C}^{in}}p_{K}(z,x_{i},t-t_{i})r_{K}(x_{i},x_{i-1},t_{i}-t_{i-1})...r_{K}(x_{1},w,t_1)\prod_{j=1}^{i}dvol_{x_j}.$$
where $t\Delta_{i}=\{(t_{1},t_{2},...,t_{i})\in \mathbb{R}^{i}~|~0\leq t_{1}\leq t_{2}\leq...\leq t_{i}\leq t\}$.

We prove that if $K$ is large enough, this integral is convergent and that $p_{K}^{i}(z,w,t)$ is differentiable to some order depending on $K$.
Define
$$r_{K}^{i+1}(z,w,t)=\int_{t\Delta_{i}}dt_{1}dt_{2}...dt_{i}\int_{\mathbb{C}^{in}}r_{K}(z,x_{i},t-t_i)r_{K}(x_{i},x_{i-1},t_{i}-t_{i-1})...r_{K}(x_{1},w,t_1)\prod_{j=1}^{i}dvol_{x_j},$$
then we have
\bl If $q_{1}-q_{n}<\frac{1}{3}$, $l_{0}\in\mathbb{N}, ~\frac{K\delta}{3}-\frac{2}{3(1-q_1)}-n-l_{0}-\frac{l_{0}q_1}{2-2q_1}>0$, then $r_{K}^{i+1}$ is of class $C^{l_0}$ with respect to $z,\bar{z}$, and
\begin{flalign*}
\|r_{K}^{i+1}\|_{0}\leq~& t^{[\frac{K\delta}{3}-\frac{2}{3(1-q_1)}](i+1)-n}[H(t)]^{i+1}\frac{t^{i}}{i!},\\
\|r_{K}^{i+1}\|_{l_0}\leq~& t^{[\frac{K\delta}{3}-\frac{2}{3(1-q_1)}](i+1)-n-l_{0}-\frac{l_{0}q_1}{2-2q_1}}H_{l_{0},k}(t)[H(t)]^{i}\frac{t^{i}}{i!}.
\end{flalign*}
\el

\bl Assume $\frac{K\delta}{3}-\frac{2}{3(1-q_1)}-n-l_{0}-\frac{l_{0}q_1}{2-2q_1}>0$.

\noindent(1)~The kernel $p_{K}^{i}$ is of class $C^{l_0}$ with respect to $z,\bar{z}$, and there exists a constant $\widetilde{C}$ such that
\begin{flalign*}
\|p_{K}^{i}\|_{0}\leq~& \widetilde{C}t^{[\frac{K\delta}{3}-\frac{2}{3(1-q_1)}]i-n}H(t)^{i}\frac{t^{i-1}}{(i-1)!}, \quad \text{for } i\geq 1,\\
\|p_{K}^{i}\|_{l_0}\leq~& \widetilde{C}t^{[\frac{K\delta}{3}-\frac{2}{3(1-q_1)}]i-n-l_{0}-\frac{l_{0}q_1}{2-2q_1}}H(t)^{i}\frac{t^{i-1}}{(i-1)!}, \quad \text{for } i\geq 1
\end{flalign*}
(2)~$p_{K}^{i}(z,w,t)$ is of class $C^{1}$ with respect to t, and
$$(\p_{t}+\Delta_{f}^{0})p_{K}^{i}(z,w,t)=r_{K}^{i+1}(z,w,t)+r_{K}^{i}(z,w,t).$$
\el

Analogously, we can consider the derivatives with respect to $w$. Since $z$ and $w$ are almost on the same footing, the estimate results of $w$-derivatives are the same.

Proceeding as the proof in \cite{1992Heat}, we get Theorem \ref{kernel} in the 0-form and $2n$-form case.
\bt Assume that the kernel $p_{K}(z,w,t)$ satisfies the conditions of Proposition \ref{estimater}. For any $l_{0}\in\mathbb{N}$, such that $\frac{K\delta}{3}-\frac{2}{3(1-q_1)}-n-l_{0}-\frac{l_{0}q_1}{2-2q_1}>0$, the series
$$p(z,w,t)=\sum_{i=0}^{\infty}p_{K}^{i}(z,w,t)$$
converges in the $\|\cdot\|_{l_0}$-norm over $\C^{n}\times\C^{n}$, defines a $C^{1}$ map from $R_{+}$ to $C^{l_0}(\C^{n}\times\C^{n})$ and
$$(\p_{t}+\Delta_{f_0}^{0})p(z,w,t)=0.$$
The kernel $p(z,w,t)$ is the heat kernel function for $\Delta_{f_0}^{0}$.
\et

For $1\leq k\leq 2n-1$, there is a $B$-term in the matrix differential operator $\Delta_{f_0}^{k}$. The construction is almost the same, since we can regard the $B$-term as a perturbation.
In this case, we use the same notation, and $\mathcal{E}_{0}=\frac{1}{(2\pi t)^{n}}\exp{(-\frac{|z-w|^2}{2t})},\m{E}_{1}=\exp{(-tg)}$ as well. We have
\begin{flalign*}
&\text{Coeff}(t^{-1})=r\p_{r}U_{0},\\
&\text{Coeff}(t^0)=r\p_{r}U_{1}+U_{1}-2\p_{z_{\nu}}\p_{\bar{z}_{\nu}}U_{0}+BU_{0},\\
&\text{Coeff}(t^1)=r\p_{r}U_{2}+2U_{2}-2\p_{z_{\nu}}\p_{\bar{z}_{\nu}}U_{1}+BU_{1}+2(g_{z_{\nu}\bar{z}_{\nu}}U_{0}+g_{z_{\nu}}\p_{\bar{z}_{\nu}}U_{0}+g_{\bar{z}_{\nu}}\p_{z_{\nu}}U_{0}),\\
&\text{Coeff}(t^a)=r\p_{r}U_{a+1}+(a+1)U_{a+1}-2\p_{z_{\nu}}\p_{\bar{z}_{\nu}}U_{a}+BU_{a}+2(g_{z_{\nu}\bar{z}_{\nu}}U_{a-1}+g_{z_{\nu}}\p_{\bar{z}_{\nu}}U_{a-1}\\
&\quad \quad \quad \quad \quad +g_{\bar{z}_{\nu}}\p_{z_{\nu}}U_{a-1})-2g_{z_{\nu}}g_{\bar{z}_{\nu}}U_{a-2} \quad \text{for }a\geq2.
\end{flalign*}

\noindent Note that $L_{f_0}=-(h^{\bar{\mu}\nu}\p_{\nu}{f_0}_{l}\iota_{\p_{\bar{\mu}}}dz^{l}\wedge+\overline{h^{\bar{\mu}\nu}\p_{\nu}{f_0}_{l}\iota_{\p_{\bar{\mu}}}dz^{l}\wedge})$, so we have $\p_{z_{\mu}}\p_{\bar{z}_{\mu}}B=0$. Then each equation Coeff$(t^a)=0$, $U_{a}$ is solved as follows:
\begin{equation}\label{eq:Unew}
\begin{aligned}
U_{0}(z,w)~=~&I \text{  is the identity matrix,}\\
U_{1}(z,w)~=~&-\int_{0}^{1}B(\tau(z-w)+w)d\tau,\\
U_{2}(z,w)~=~&\frac{1}{r^{2}}\int_{0}^{r}s\left(-2g_{z_{\nu}\bar{z}_{\nu}}+B\int_{0}^{1}B(\tau(z-w)+w)d\tau\right)ds\\
      =~&\int_{0}^{1}d\lambda\int_{0}^{\lambda}(-2\p_{z_{\nu}}\p_{\bar{z}_{\nu}}|\p f|^{2})(\tau(z-w)+w)\tau d\tau\\
        &+\int_{0}^{1}d\lambda\int_{0}^{\lambda}B(\tau(z-w)+w)B(\lambda(z-w)+w)d\tau,\\
U_{a}(z,w)~=~&U_{a}(z-w,w)\\
=~&\int_{0}^{1}\left\{2(\p_{\nu}\p_{\bar{\nu}}U_{a-1})(\tau(z-w),w)-BU_{a-1}(\tau(z-w),w)\right.\\
  &-2[(\p_{\nu}\p_{\bar{\nu}}g)(\tau(z-w),w)U_{a-2}(\tau(z-w),w)\\
 ~&+(\p_{\nu}g)(\tau(z-w),w)(\p_{\bar{\nu}}U_{a-2})(\tau(z-w),w)\\
  &+(\p_{\bar{\nu}}g)(\tau(z-w),w)(\p_{\nu}U_{a-2})(\tau(z-w),w)]\\
 ~&\left.+2(g_{\nu}g_{\bar{\nu}})(\tau(z-w),w)U_{a-3}(\tau(z-w),w)\right\}\tau^{a-1}d\tau,\quad a\geq3.
\end{aligned}
\end{equation}

\noindent The remainder $r_{K}(z,w,t)$ becomes
\begin{flalign*}
r_{K}(z,w,t)=~&(\p_{t}+B+\Delta_{f})p_{K}(z,w,t)\\
              =~&\m{E}_{0}\m{E}_{1}\left[(-2\p_{z_{\nu}}\p_{\bar{z}_{\nu}}U_{K}+BU_{K}+2g_{z_{\nu}\bar{z}_{\nu}}U_{K-1}+2g_{z_{\nu}}\p_{\bar{z}_{\nu}}U_{K-1}+2g_{\bar{z}_{\nu}}\p_{z_{\nu}}U_{K-1}-2g_{z_{\nu}}g_{\bar{z}_{\nu}}U_{K-2})t^{K}\right.\\
                &\left.+2(g_{z_{\nu}\bar{z}_{\nu}}U_{K}+g_{z_{\nu}}\p_{\bar{z}_{\nu}}U_{K}+g_{\bar{z}_{\nu}}\p_{z_{\nu}}U_{K}-g_{z_{\nu}}g_{\bar{z}_{\nu}}U_{K-1})t^{K+1}-2g_{z_{\nu}}g_{\bar{z}_{\nu}}U_{K}t^{K+2}\right]\\
              =~:&~\m{E}_{0}\m{E}_{1}\widetilde{U}.
\end{flalign*}

\noindent We want to extract a positive power of $t^{K}$ in $\widetilde{U}$ after the rescaling, so we only need to consider the maximal weight term in $\widetilde{U}$. Define $\w(B)= \max\{\w(B_{ij})\}$, then $\w(B)=1-2q_{n}$. Inductively, for $a\geq 3$, the maximal weight contribution in recursion formula \eqref{eq:Unew} of $U_a$ comes from the term $\sum_{\mu=1}^{n}g_{z_{\mu}}g_{\bar{z}_{\mu}}U_{a-3}$. Therefore, there is at most one $B$-term contribution in the maximal weight in $\widetilde{U}$. Essentially, there is no difference from the previous case. Explicitly, denote $\w(U_a)$ the biggest weight in $U_a$. Using \eqref{eq:Unew}, we have

\bl\label{wtdiscussion} For $j\geq 1$, $\w(U_{3j})=(4-6q_n)j,~ \w(U_{3j+1})=1-2q_{n}+(4-6q_n)j,~ \w(U_{3j+2})=2-4q_{n}+(4-6q_n)j.$
\el

Then when $q_{1}-q_{n}<\frac{1}{3}$, the remainder $r_{K}(z,w,t)$ satisfies
$$|r_{K}(z,w,t)|\leq t^{K-\frac{1-2q_n}{1-q_1}-\frac{2-3q_n}{1-q_1}(\frac{K-2}{3}+1)}H(t)= t^{\frac{K}{3}\delta-\frac{5-9q_n}{3(1-q_1)}}H(t).$$

Therefore, we get the existence and uniqueness of the heat kernel funtion for $\Delta_{f_0}$, i.e. Theorem \ref{kernel}.\\

Now let us consider the deformed case. Let $f(z;u)=f_0(z)+\sum_{i=1}^su^i\phi_i$ be the relevant or the marginal deformation satisfying the weight condition $(\star)$.

\bp[Proof idea of Theorem \ref{deformedkernel}] In the sequel, we view $\{u^i\}_{i=1}^s$ as parameters.
\begin{enumerate}
  \item If $f$ is the marginal deformation, then we can define the parametrix $p_K(z,w,t;u)$ as
  $$p_K(z,w,t;u)=\m{E}_0\m{E}_1\sum_{a=0}^KU_a(z,w;u)t^a,$$
  where $\m{E}_0$ is the same as before, $\m{E}_1=\exp[-tg(z,w;u)]$, with
  $$g(z,w;u)=2\int_0^1\left|\p f\right|^2(\tau(z-w)+w)d\tau.$$
  Then we can proceed as before to obtain the heat kernel function.
  \item If $f$ is the relevant deformation with $\w(\phi_i)<1-2(q_1-q_n)$, then we can define the parametrix $p_K(z,w,t;u)$ as
  $$p_K(z,w,t;u)=\m{E}_0\m{E}_1\sum_{a=0}^KU_a(z,w;u)t^a,$$
  where $\m{E}_0$ is the same as before, $\m{E}_1=\exp[-tg(z,w)]$, with
  $$g(z,w)=2\int_0^1\left|\p f_0\right|^2(\tau(z-w)+w)d\tau.$$
  i.e. we add the term $(|\p f|^2-|\p f_0|^2)I=2\Re(\sum_{i=1}^su^i\p\phi_i\overline{\p f_0})+|\sum_{i=1}^su^i\p\phi_i|^2$ into the $B$-term. Then the weight of new $B$-term is less than $2(1-q_1)$, furthermore, the maximal weight in each $U_a$ may come from the two cases
  \begin{itemize}
  	\item the iteration of the new $B$ term, which gives the weight less than $2a(1-q_1)$;
  	\item the iteration of the term $\sum_{\mu=1}^ng_{z_{\nu}}g_{\bar{z}_{\nu}}$, which gives the weight as Lemma \ref{wtdiscussion} writes.
  \end{itemize}
  We can also extract a positive power of the remainder in both cases. Then we are done.
\end{enumerate}
\ep

\end{appendices}


\begin{thebibliography}{0}

\bibitem{Bershadsky1993Holomorphic}
Bershadsky, M. and Cecotti, S. and Ooguri, H. and Vafa, C.
``Holomorphic Anomalies in Topological Field Theories."
\textit{Nuclear Physics B}, 405, no. 2-3(1993): 279-304.


\bibitem{Bershadsky1994Kodaira}
Bershadsky, M. and Cecotti, S. and Ooguri, H. and Vafa, C.
`` Kodaira-Spencer Theory of Gravity and Exact Results for Quantum String Amplitudes."
\textit{Communications in Mathematical Physics}, 165, no. 2(1994): 311-427.


\bibitem{Bismut1992Higher}
Bismut, Jean Michel and Kai, K\"{a}hler
``Higher Analytic Torsion Forms for Direct Images and Anomaly Formulas."
\textit{Journal of Algebraic Geometry}, 1, no.4(1992).


\bibitem{Cecotti1992A}
Cecotti, S. and Fendley, P. and Intriligator, K. and Vafa, C.
``A New Supersymmetric Index."
\textit{Nuclear Physics B}, 386, no. 2(1992): 405-452.


\bibitem{Cecotti1991Topological}
Cecotti, S. and Vafa, C.
``Topological-anti-topological Fusion.''
\textit{Nuclear Physics B} 367, no. 367 (1991): 359-461.


\bibitem{Cecotti1993Ising}
Cecotti, S. and Vafa, C.
``Ising Model and N =2 Supersymmetric Theories."
\textit{Communications in Mathematical Physics}, 157, no. 1(1993): 139-178.


\bibitem{Coates2015A}
Coates, Tom and Iritani, Hiroshi
``A Fock Sheaf For Givental Quantization."
\textit{arXiv: 1411.7039}.


\bibitem{Cognola2006Heat}
Cognola, G. and Elizalde, E. and Zerbini, S.
``Heat-kernel Expansion on Noncompact Domains and a Generalized Zeta-function Regularization Procedure."
\textit{Journal of Mathematical Physics}, 47, no. 8(2006): 3516-083516.


\bibitem{Costello2012Quantum}
Costello, K. and Li, S.
``Quantum BCOV Theory on Calabi-Yau Manifolds and the Higher Genus B-model." 2012.
\textit{arXiv:1201.4501}.


\bibitem{Fan2011Schr}
Fan, H.
``Schr\"odinger equations, deformation theory and $tt^*$-geometry."
\textit{arXiv:1107.1290}.


\bibitem{Fan2016Torsion}
Fan, H. and Fang, H.
``Torsion type invariants of singularities."
\textit{arXiv:1603.06530v1}.

\bibitem{Fang2008Asymptotic}
Fang H, Lu Z, Yoshikawa K.
``Asymptotic Behavior of the BCOV Torsion of Calabi-Yau Moduli."
\textit{Journal of Differential Geometry}, 80, 2008, 175-259.


\bibitem{SouleAnalytic}
J-M.Bismut, H.Gillet and C.Soule
``Analytic Torsion and Holomorphic Determinant Bundles. I: Bott-Chern Forms and Analytic Torsion. II: Direct Images and Bott-Chern Forms. III: Quillen Metrics on Holomorphic Determinant Bundles."
\textit{Communications in Mathematical Physics}, 115, no. 1(1988): 49-78; 79-126; 301-351.


\bibitem{Li2013Primitive}
Li, C. and Li, S. and Saito, K.
``Primitive Forms via Polyvector Fields."
\textit{arXiv: 1311.1659}.


\bibitem{LiquantumB}
Li S.
``On the Quantum Theory of Landau-Ginzburg B-model."
\textit{in preparation}.


\bibitem{1992Heat}
Nicole Berline, Ezra Getzler and Michele Vergne
``Heat kernel and Dirac operators."(1992).


\bibitem{Ray1973Analytic}
Ray, D. B. and Singer, I. M.
``Analytic Torsion for Complex Manifolds."
\textit{Annals of Mathematics}, 98, N0. 1(1973): 154-177.




\bibitem{Saito1981Primitive}
Saito, K.
``Primitive Forms for a Universal Unfolding of a Function with an Isolated Critical Point."
\textit{Faculty of Science, The University of Tokyo}, no 3(1982): 775-792.


\bibitem{Saito1983The}
Saito, K.
``The Higher Residue Pairings $\mathcal{K}_F$ for a Family of Hypersurface Singular Points."
\textit{Proceedings of Symposia in pure mathematics}, (1983): 441-463.





\end{thebibliography}
\end{document}